\documentclass[12pt]{article}
\usepackage{amsmath,amssymb}
\usepackage{ifpdf}
\makeatletter

    \@addtoreset{equation}{section}
\makeatother
\newsavebox{\boxa}
\newlength{\bw}
\ifpdf
  \usepackage[pdftex]{graphicx}
\else
  \usepackage[dvipdfm]{graphicx}
\fi

\begin{document}

\thispagestyle{empty}

\vspace*{3cm}

\begin{center}
 {\LARGE {The gravity duals of $SO/USp$ superconformal\\[2mm] quivers}}
\vskip2cm
{\large 
{Takahiro Nishinaka\footnote{nishinak@post.kek.jp}
}
}
\vskip.5cm
{\it Center for Quantum Spacetime (CQUeST), Sogang University
\\
Seoul 121-742, Korea}
\end{center}

\vskip1cm
\begin{abstract}
We study the gravity duals of $SO/USp$ superconformal quiver gauge theories realized by wrapping M5-branes on a Riemann surface (``G-curve'') together with a $\mathbb{Z}_2$-quotient. When the G-curve has no punctures, the gravity solutions are classified by the genus $g$ of the G-curve and the torsion part of the four-form flux $G_4$. We also find that there is an interesting relation between anomaly contributions from two mysterious theories: $T_{SO(2N)}$ theory with $SO(2N)^3$ flavor symmetry and $\widetilde{T}_{SO(2N)}$ theory with $SO(2N)\times USp(2N-2)^2$ flavor symmetry. The dual gravity solutions for various $SO/USp$-type tails are also studied.
\end{abstract}


\newpage
\tableofcontents

\section{Introduction}

Recently, there has been remarkable progress in the study of $d=4,\mathcal{N}=2$ superconformal field theories which are realized by wrapping M5-branes on a punctured Riemann surface. This was triggered by the work \cite{Gaiotto:2009we} where a large class of $SU(N)$ superconformal quiver gauge theories are constructed from wrapped M5-branes. The Riemann surface wrapped by the M5-branes is called the ``G-curve,'' whose complex structure moduli are identified with the marginal couplings of the theories. The flavor symmetry of the theory is associated with the punctures on the G-curve.
The detailed study of S-duality for such theories leads to the still mysterious $T_N$-theory which has $SU(N)^3$ flavor symmetry and no marginal couplings.

The gravity duals of such $SU(N)$ superconformal quivers were studied in \cite{Gaiotto:2009gz}, by using the knowledge of the half-BPS solutions of eleven-dimensional gravity \cite{Maldacena:2000mw}. The dual geometry is of the form $AdS_5\times X_{6}$ where $X_6$ involves the same Riemann surface and has $SU(2)\times U(1)$ symmetry. It was shown that the holographic calculation of the central charge is consistent with the conformal anomalies of $T_{N}$-theory which are predicted by S-duality invariance. The dual gravity solutions associated with the various $SU$-type tails of quivers were also studied in \cite{Gaiotto:2009gz}, which are given in terms of the solutions of a three-dimensional Toda equation.

On the other hand, the generalization of \cite{Gaiotto:2009we} to the $SO/USp$-type quiver gauge theories are studied in \cite{Tachikawa:2009rb}. In the type IIA construction of $d=4,\mathcal{N}=2$ theories, $SO/USp$-type gauge groups in four dimensions require O4-planes in the D4/NS5/D6-system. Correspondingly, its M-theory lift involves a $\mathbb{Z}_2$-quotient. The space of marginal couplings are again identified with the moduli space of the G-curve, but there are now two different counterparts of $T_{N}$-theory; one has $SO(2N)^3$ flavor symmetry and the other has $SO(2N)\times USp(2N-2)^2$ flavor symmetry. The various $SO/USp$-type punctures and tails were also classified in \cite{Tachikawa:2009rb}. For some works on these theories and related topics, see e.g. \cite{Benini:2010uu, Tachikawa:2010vg, Nanopoulos:2010bv, Hollands:2010xa, Chacaltana:2011ze, Hollands:2011zc}.

In this paper, we study the gravity duals of such $SO/USp$ superconformal quivers, that is, the $SO/USp$ counterpart of \cite{Gaiotto:2009gz}. Since the two counterparts of the $T_{N}$-theory are still mysterious in the field theory side, the corresponding gravity duals are worth studying. The main difference from the $SU(N)$ quivers is the $\mathbb{Z}_2$-quotient in the bulk. We first identify the dual gravity of $SO/USp$ quivers whose G-curve is a Riemann surface of genus $g$ without punctures. By using the knowledge of the M-theory lifts of O4-planes, we can identify the correct $\mathbb{Z}_2$-quotient. A holographic calculation in the resulting geometry gives the correct conformal anomalies of the corresponding quiver gauge theory.

What is interesting here is that we can now attach two different gauge groups, namely $SO$ and $USp$, to the handles of the G-curve. Correspondingly, there is a class of theories which share the same G-curve without punctures. We show that such theories have the same conformal anomalies, which implies that their gravity duals share the same metric. Such dual gravities of the $SO/USp$ quivers with the same G-curve are further classified by the torsion part of the four-form flux which is associated to the ``$B$-cycles'' of the G-curve. 

We also discuss the dual gravities associated with the various $SO/USp$-type tails of quivers. The crucial point is again how to identify the $\mathbb{Z}_2$-quotient in the bulk. In particular, the quotient should be consistent with the fact that crossing a D6-brane in the type IIA configuration replaces $\mathbb{RP}^4\times S^1$ with $(S^4\times S^1)/\mathbb{Z}_2$, and vice versa, in the eleven-dimensional near horizon geometry. We identify such a proper $\mathbb{Z}_2$-quotient in the bulk.

This paper is organized as follows. In section \ref{sec:SO/USp_quivers}, we first review the type IIA brane construction of the $SO/USp$ superconformal quivers and their M-theory lifts. We also review the two mysterious theories with $SO(2N)^3$ or $SO(2N)\times USp(2N-2)^2$ flavor symmetry and no marginal couplings. In section \ref{sec:gravity1} we consider the gravity duals of $SO/USp$ quivers whose G-curve is a Riemann surface without punctures, and in section \ref{sec:gravity2} we focus on the dual gravity solutions of various $SO/USp$-type punctures.

Throughout this paper, we follow the notation of \cite{Tachikawa:2009rb} unless otherwise stated. Note also that we always count the number of M5-branes on a $\mathbb{Z}_2$-orbifold in the covering space.

\section{$SO/USp$ quivers from wrapped M5-branes}
\label{sec:SO/USp_quivers}

In this section, we review the M-theory construction of $SO/USp$ superconformal quivers.
In subsection \ref{subsec:O4-planes}, we first recall the classification of O4-planes and review the type IIA constructions of $SO/USp$ superconformal quivers. Their M-theory lifts are described in \ref{subsec:M-theory_lift}. In \ref{subsec:S-duality}, we briefly review the work \cite{Tachikawa:2009rb} to explain that the S-duality invariance of the quiver gauge theories leads to $T_{SO(2N)}$ and $\widetilde{T}_{SO(2N)}$ theories; the former has $SO(2N)^3$ flavor symmetry while the latter has $SO(2N)\times USp(2N-2)^2$ flavor symmetry.

\subsection{O4-planes and $SO/USp$ superconformal quivers}
\label{subsec:O4-planes}

As was shown in \cite{Hori:1998iv}, the orientifold four plane has four different types, which are classified by the topology of NSNS B-field and RR $U(1)$ gauge field:
\begin{eqnarray}
\vartheta \equiv \frac{1}{2\pi}\int_{\mathbb{RP}^2}B_2,\qquad \varphi \equiv \frac{1}{2\pi}\int_{S^1}C_1,
\label{eq:topology}
\end{eqnarray}
where $S^1$ and $\mathbb{RP}^2$ are cycles in $\mathbb{RP}^4$ which surrounds the O4-plane.
The two phases $\vartheta$ and $\varphi$ take values in $H^3(\mathbb{RP}^4,\widetilde{\mathbb{Z}})\simeq \mathbb{Z}_2$ and $H^2(\mathbb{RP}^4,\mathbb{Z})\simeq \mathbb{Z}_2$ respectively, where $\widetilde{\mathbb{Z}}$ is a sheaf of integers twisted by the orientation bundle of $\mathbb{RP}^4$.\footnote{In other words, when we go around a non-contractible loop of $\mathbb{RP}^4$, a section of $\widetilde{\mathbb{Z}}$ receives a reversed sign.} Thus, the four different O4-planes are classified by $(\vartheta,\varphi)\in \mathbb{Z}_2\times \mathbb{Z}_2$. Their properties are summarized in table \ref{table:O4-1}.
\begin{table}[h]
\begin{center}
\begin{tabular}{|c|c|c|}\hline
O4 & $(\vartheta,\varphi)$ & Gauge theory in four dimensions\\
\hline
 O4$^-$ & $\left(0,0\right)$ & 
       $SO(2N)$ gauge theory\\[1mm]
O4$^+$ & $\left(1,0\right)$ &
        $USp(2N)$ gauge theory\\[1mm]
O4$^{0}$ & $\left(0,1\right)$&
       $SO(2N+1)$ gauge theory\\[1mm]
$\widetilde{\rm O4}^{+}$ & $\left(1,1\right)$ &
       $USp(2N)$ gauge theory\\[.5mm]
\hline
\end{tabular}
\caption{There are four types of O4-planes which are classified by two discrete phases $(\vartheta,\varphi) \in \mathbb{Z}_2\times \mathbb{Z}_2$. The gauge group which arises in four dimensions depends on the type of O4-planes.}
\label{table:O4-1}
\end{center}
\end{table}

\begin{figure}
\begin{center}
\includegraphics[width=13cm]{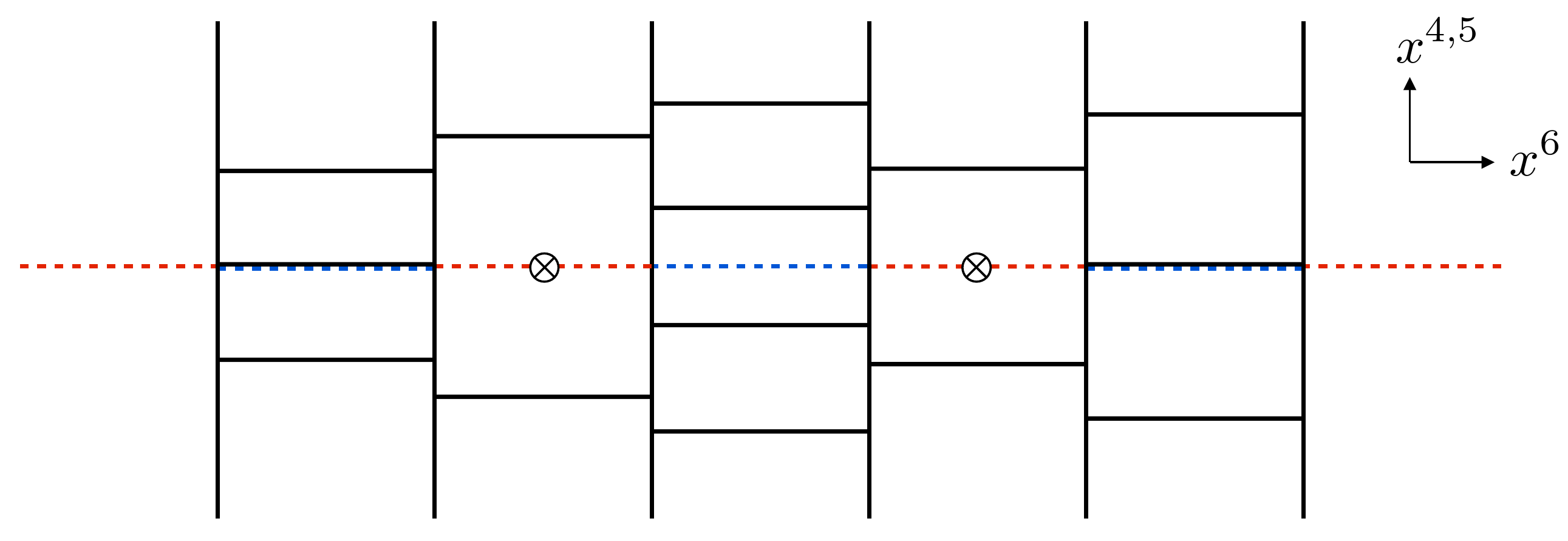}
\caption{An example of the type IIA construction of the $SO/USp$ superconformal quiver gauge theories in four dimensions. The solid vertical lines denote NS5-branes while the solid horizontal lines show D4-branes. The dotted lines express O4-planes whose colors are determined by the value of $\vartheta$.   This example realizes $SO(3)^2\times USp(2)^2 \times SO(4)$ gauge theory with two fundamentals in four dimensions.}
\label{fig:quiver}
\end{center}
\end{figure}

By using D4, O4 and NS5-branes, we can construct a type IIA realization of $d=4,\mathcal{N}=2$ linear quiver gauge theories with $SO/USp$ gauge groups \cite{Evans:1997hk, Landsteiner:1997vd, Brandhuber:1997cc, deBoer:1998by, Hori:1998iv, Gimon:1998be, Tachikawa:2009rb}, which is the $SO/USp$ generalization of the brane construction of $SU$-type quivers \cite{Witten:1997sc}. A typical brane configuration is depicted in figure \ref{fig:quiver}.
D4-branes are located at $x^{7,8,9}=0$ and localized in the $(x^4, x^5)$-plane, while NS5-branes are located at $x^{7,8,9}=0$ and localized in $x^6$-direction. We also have an O4-plane at $x^{4,5,7,8,9}=0$, which realizes the $SO/USp$ gauge groups in four dimensions. Note that the O4-plane has intersections with the NS5-branes along $x^6$-axis. Since an NS5-brane carries one magnetic charge for NSNS two-form, crossing a NS5-brane shifts $\vartheta$ by one unit. This replaces O4$^{-}$ with O4$^{+}$ as well as O4$^0$ with $\widetilde{\rm O4}^{+}$, and vice versa.

We can also introduce D6-branes at some definite values of $x^{4,5,6}$. Such D6-branes give additional fundamental matters in the low energy gauge theory. In this paper, we assume that all the D6-branes are at $x^{4,5}=0$ just for simplicity, which physically means that the fundamental matters in four dimensions are massless. In this situation, the O4-plane is also divided by the D6-branes. Since a D6-brane carries one unit of magnetic charge for RR $U(1)$ gauge field, crossing D6-brane shifts $\varphi$ by one unit, which replaces O4$^-$ with O4$^0$ as well as O4$^{+}$ with $\widetilde{\rm O4}^+$, and vice versa.

When the brane configuration has $(n+1)$ NS5-branes, the low-energy gauge theory in four dimensions includes $n$ gauge groups. We here define $d_i$ for $i=1,2,\cdots,n$ so that the gauge group associated with an interval between $i$-th and $(i+1)$-th NS5-branes is $SO(d_i)$ or $USp(d_i-2)$. Then, the conformal symmetry in four dimensions implies
\begin{eqnarray}
 k_i = 2d_i - d_{i-1}-d_{i+1},
\label{eq:superconformal_SO/USp}
\end{eqnarray}
where $k_i$ is the number of D6-branes between $i$-th and $(i+1)$-th NS5-branes.
Then, for a general $SO/USp$ superconformal linear quiver, we have
\begin{eqnarray}
 d_1<\cdots<d_{\ell}= d_{\ell+1}=d_{r}>d_{r+1}>d_n,
\end{eqnarray}
where $\ell$ and $r$ are some positive integers satisfying $\ell\leq r$. We call the left $\ell$ and the right $(n-r+1)$ gauge groups the left and right ``{\em tails}.''
For example, a brane configuration in figure \ref{fig:quiver} realizes $SO(3)^2\times USp(2)^2 \times SO(4)$ superconformal gauge theory with two fundamentals in four dimensions.

\subsection{M-theory lift}
\label{subsec:M-theory_lift}

We now consider the M-theory lifts of the above type IIA configurations.
In \cite{Hori:1998iv}, the M-theory lifts of the O4-planes are identified as in table \ref{table:O4-2}.
\begin{table}
\begin{center}
\begin{tabular}{|l|l|p{11cm}|}\hline
O4 & $(\vartheta,\varphi)$ & \hspace{4cm} M-theory lift\\
\hline\hline
O4$^-$ & $(0,0)$ & M-theory on $\mathbb{R}^5\times \mathbb{R}^5/\mathbb{Z}_2 \times S^1$\\
\hline
O4$^+$ & $(1,0)$ & M-theory on $\mathbb{R}^5\times \mathbb{R}^5/\mathbb{Z}_2 \times S^1$. A pair of M5-branes are localized at the $\mathbb{Z}_2$-fixed plane.\\
\hline
O4$^0$ & $(0,1)$ & M-theory on $\mathbb{R}^5\times (\mathbb{R}^5\times S^1)/\mathbb{Z}_2$.\\
\hline
 $\widetilde{\rm O4}^{+}$ & $(1,1)$ & M-theory on $\mathbb{R}^5\times (\mathbb{R}^5\times S^1)/\mathbb{Z}_2$. A single M5-brane is localized at the $\mathbb{Z}_2$-invariant cylinder.\\
\hline
\end{tabular}
\caption{The M-theory lifts of O4-planes.}
\label{table:O4-2}
\end{center}
\end{table}
Here, $\mathbb{R}^5$ acted by $\mathbb{Z}_2$ is transverse to the O4-planes, and spanned by $x^{4,5,7,8,9}$. The non-trivial $\mathbb{Z}_2$-action on $S^1$ is a half-period shift of M-theory direction $x^{10}$. This classification implies that the $\mathbb{Z}_2$-action on $S^1$ is trivial or non-trivial depending on the value of $\varphi$. In the M-theory lift of an O4$^+$-plane, two M5-branes are ``freezing'' at the $\mathbb{Z}_2$ fixed plane and cannot move away. This is understood as a consequence of the non-vanishing torsion element of the four-form flux $G_4$ \cite{Hori:1998iv}.

\begin{figure}
\begin{center}
\includegraphics[width=15cm]{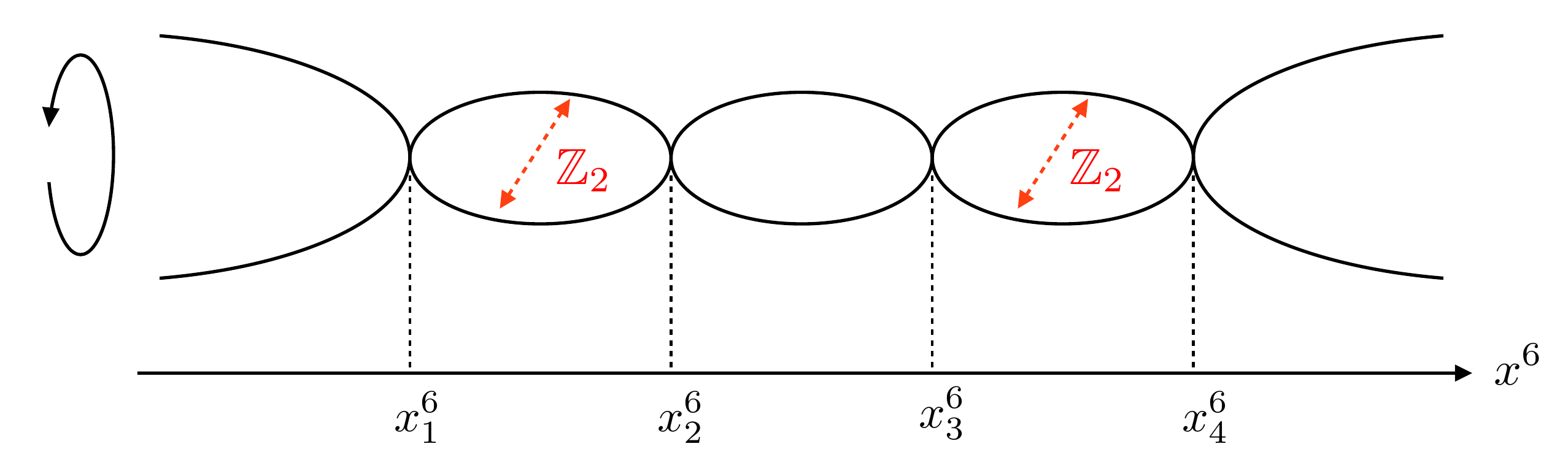}
\caption{A subspace $x^4=x^5=0$ in the multi Taub-NUT space associated with a $SO/USp$-type tail. The D6-branes are localized at $x^6=x^6_i$ in type IIA setup. The non-trivial $\mathbb{Z}_2$-quotient appears or disappears when you cross $x^6=x^6_i$.}
\label{fig:Taub-NUT}
\end{center}
\end{figure}

On the other hand, the D6-branes are lifted to a multi Taub-NUT geometry in M-theory, whose metric is given by
\begin{eqnarray}
 ds^2_{\rm TN} = \frac{V}{4}d\vec{x}^2 + \frac{V^{-1}}{4}\left(d\eta + \vec{\omega}\cdot \vec{x}\right)^2,
\end{eqnarray}
where
\begin{eqnarray}
 V = 1 + \sum_{a}\frac{1}{|\vec{x}-\vec{x}_a|},\qquad \vec{\nabla} \times \vec{\omega} = \vec{\nabla}V,
\end{eqnarray}
and $\vec{x}=(x^4,x^5,x^6)$ in our notation. The position $\vec{x}_a$ in the three dimensions expresses the location of $a$-th D6-brane, where $x^{4,5}_a = 0$ because we have assumed the vanishing fundamental masses. The subspace $x^{4,5}=0$ in the Taub-NUT space is a chain of two-cycles as shown in figure \ref{fig:Taub-NUT}, where the $S^1$-fiber degenerates at each point of $x^6=x^6_a$. When the left tail of the quiver involves $\ell$ D6-branes, we call a special $\mathbb{P}^1$ between $x^6=x^6_{\ell}$ and $x^6=x^6_{\ell+1}$ the ``middle'' $\mathbb{P}^1$.
We here see how the $\mathbb{Z}_2$-quotient, which is the lift of the O4-plane, affects these two cycles. Recall that the $\mathbb{Z}_2$-action on $x^6$ is trivial since the O4-plane is extending along $x^6$-direction. On the other hand, the $\mathbb{Z}_2$ acts on the $S^1$-fiber trivially or non-trivially, depending on $\varphi$. Since crossing the D6-brane shifts $\varphi$ by one unit, the non-trivial $\mathbb{Z}_2$-action on $S^1$ appears or disappears when one of $x^6_a$ is crossed (See figure \ref{fig:Taub-NUT}).

Finally, the D4-branes and NS5-branes in the type IIA configuration are lifted to a single M5-brane world-volume which is embedded in the multi Taub-NUT space. Since the multi Taub-NUT space is non-trivially acted by the $\mathbb{Z}_2$, the world-volume of the M5-brane should be consistent with the $\mathbb{Z}_2$-quotient.

\subsection{S-duality and $T_{SO(2N)}$ theory}
\label{subsec:S-duality}

In this subsection we briefly review the work \cite{Tachikawa:2009rb}, where the $SO/USp$ superconformal quivers are classified by a punctured Riemann surface.

In general, it is rather complicated to analyze the M5-brane world-volume embedded in the multi Taub-NUT space which is affected by the $\mathbb{Z}_2$-quotient. However, we can simplify the situation by moving the D6-branes to $x^6\to \pm \infty$, which makes the M5-brane world-volume approximately embedded in $\mathbb{R}^3\times S^1$. It is known that the four-dimensional low-energy effective theory is independent of the positions of the D6-branes in $x^6$-direction.

When we move a D6-brane in $x^6$-direction crossing some NS5-branes, we have additional D4-branes stretching between the D6-brane and the NS5-branes because of the Hanany-Witten effect \cite{Hanany:1996ie}. Let $\widetilde{d}_i$ be the number of such newly created D4-branes between the $i$-th and $(i+1)$-th NS5-branes. Since all D6-branes are attached to either the left or right tail of the quiver, we can move all the D6-branes in the left tail to $x^6\to -\infty$ as well as those in the right tail to $x^6\to +\infty$. In this case, we have $d_i+\widetilde{d}_i=2N$ for all $i=1,2,\cdots,n$, which implies that the corresponding M5-brane world-volume is roughly regarded as the $2N$-th cover of a punctured Riemann surface.\footnote{This case is called the ``balanced case'' in section 3.2.5 of \cite{Gaiotto:2009hg}.}

The punctured Riemann surface is called the G-curve \cite{Gaiotto:2009we}, and parameterized by the coordinate $t$ on the ``middle'' $\mathbb{P}^1$ in the multi Taub-NUT space.\footnote{Now, the radius of the ``middle'' $\mathbb{P}^1$ is infinitely large.} The G-curve has $(n+3)$ punctures on it, which express the locations of M5-branes transversally intersecting with $2N$ M5-branes on the G-curve. The punctures away from $t=0$ and $t=\infty$ express the $(n+1)$ NS5-branes in the type IIA limit, while the punctures at $t=0$ and $t=\infty$ describe the left and right tails of the quiver, respectively. We denote each of the $(n+1)$ former punctures by $\times$. The punctures at $t=0$ and $t=\infty$ are characterized by Young diagrams which describe the flavor symmetry associated to the left and right tails, respectively. In particular, we will later consider $SO(2N)$ and $USp(2N-2)$ flavor symmetries at the tails. The puncture associated with $SO(2N)$ flavor symmetry is denoted by $\odot$, while that for $USp(2N-2)$ flavor symmetry is expressed by $\star$. It is also important that if the tail ends with $SO(3)$ gauge group then the tail is described by the puncture $\times$ \cite{Tachikawa:2009rb}.

\begin{figure}
\begin{center}
\includegraphics[width=6cm]{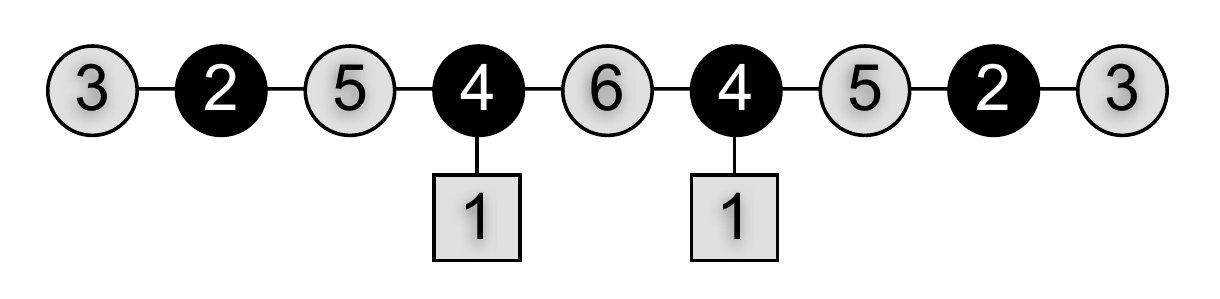}\qquad \includegraphics[width=5cm]{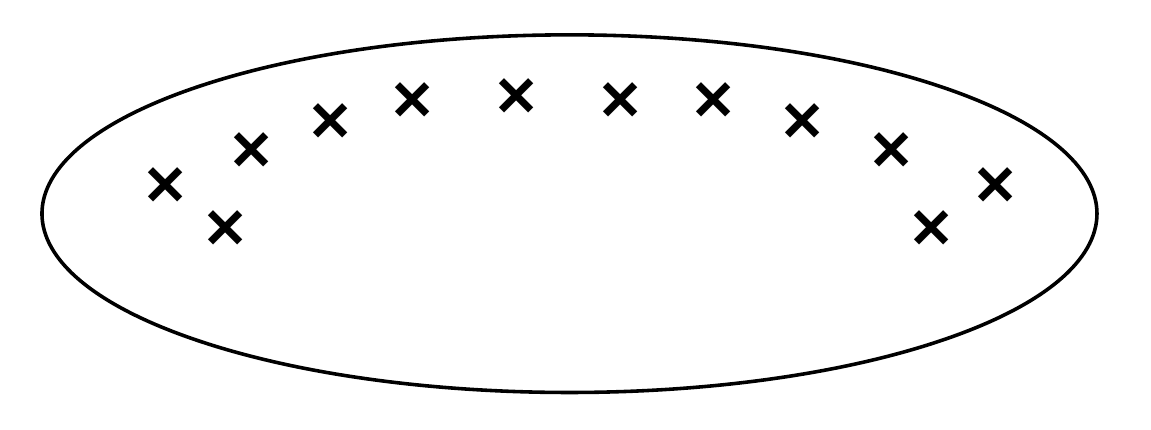}
\\[5mm]
\includegraphics[width=5cm]{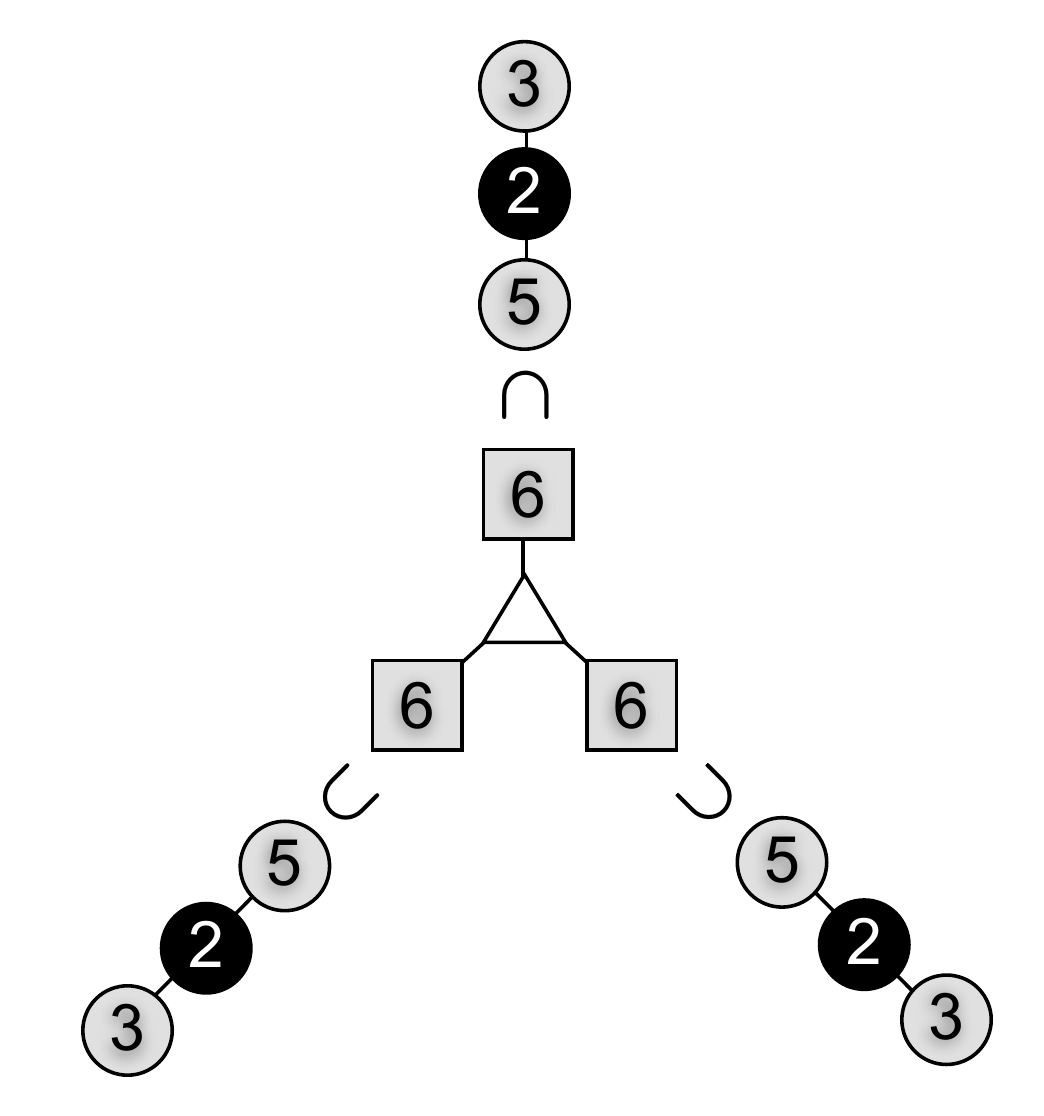}\qquad\qquad \includegraphics[width=4.5cm]{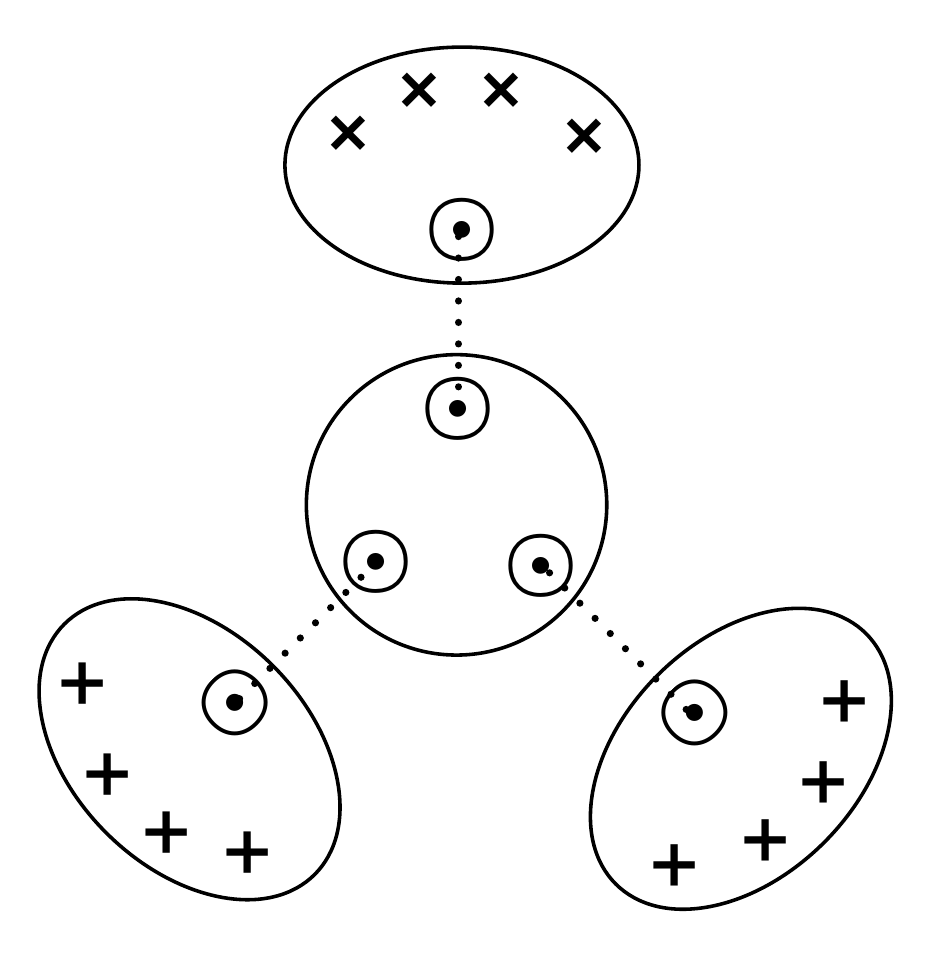}
\caption{Upper: The quiver diagram and the G-curve of a linear quiver gauge theory with the gauge group \eqref{eq:6N-9} for $N=3$. In the quiver, the grey and black circles with $n$ inside stands for $SO(n)$ and $USp(n)$ gauge groups respectively, while the number inside a box represents the flavor symmetry.\,\, Lower: The upper G-curve can be deformed as in the lower picture, making three necks associated with $\odot$. The resulting curve has a weakly coupled description which involves $T_{SO(6)}$ theory and three tails of $SO(5)\times USp(2)\times SO(3)$.}
\label{fig:T_SO}
\end{center}
\end{figure}

The parameter space of marginal coupling constants in the four-dimensional gauge theory is identified with the moduli space of the G-curve. There are some points in the moduli space where the theory has a weekly coupled description, which are related to each other by S-duality. Let us consider a linear quiver theory with $6N-9$ gauge group
\begin{eqnarray}
&&SO(3) \times USp(2) \times \cdots \times USp(2N-4) \times SO(2N-1)\times
\nonumber\\
&&\qquad\qquad USp(2N-2)\times SO(2N) \times \cdots \times SO(2N) \times USp(2N-2)\times
\nonumber\\
&&\qquad\qquad\qquad\qquad SO(2N-1) \times USp(2N-4) \times \cdots \times USp(2) \times SO(3),\qquad
\label{eq:6N-9}
\end{eqnarray}
whose G-curve is a sphere with $6N-6$ punctures of $\times$ (See figure \ref{fig:T_SO}). We can deform the G-curve so that it has three tails with $2N-2$ punctures for each. Each of the three tails describes a linear quiver of gauge group
\begin{eqnarray}
 SO(2N-1) \times USp(2N-4) \times \cdots \times USp(2) \times SO(3),
\label{eq:tail_SO}
\end{eqnarray}
 and the tails are connected with a sphere with three punctures of $\odot$.
Here, a $SO(2N-1)$ gauge group in each tail is gauging the subgroup of $SO(2N)$ flavor symmetry associated with $\odot$ on the sphere. The sphere with three $\odot$ expresses a theory with $SO(2N)^3$ flavor symmetry and no marginal gauge coupling. This theory is denoted by $T_{SO(2N)}$ in \cite{Tachikawa:2009rb}.\footnote{This is the $SO(2N)$ counterpart of $T_{N}$ theory studied in \cite{Gaiotto:2009we}.} We can determine the anomalies $a$ and $c$ of $T_{SO(2N)}$ theory so that the total anomaly of the quiver is the same as that of the original linear quiver with gauge group \eqref{eq:6N-9}. Following \cite{Gaiotto:2009gz}, we define $n_v$ and $n_h$ for $T_{SO(2N)}$ theory so that its conformal anomalies $a$ and $c$ are written as
\begin{eqnarray}
a = \frac{5n_v+n_h}{24},\qquad c= \frac{2n_v+n_h}{12}.
\label{eq:ac-nvnh}
\end{eqnarray}
The explicit expression for $n_v$ and $n_h$ of $T_{SO(2N)}$ theory was calculated in \cite{Tachikawa:2009rb} as
\begin{eqnarray}
n_{v}(T_{SO(2N)}) = \frac{8N^3}{3} -7N^2 + \frac{10N}{3}, \qquad n_{h}(T_{SO(2N)}) =\frac{8N^3}{3} - 4N^2 + \frac{4N}{3}.
\label{eq:anomaly_TSO1}
\end{eqnarray}

\begin{figure}
\begin{center}
\includegraphics[width=7cm]{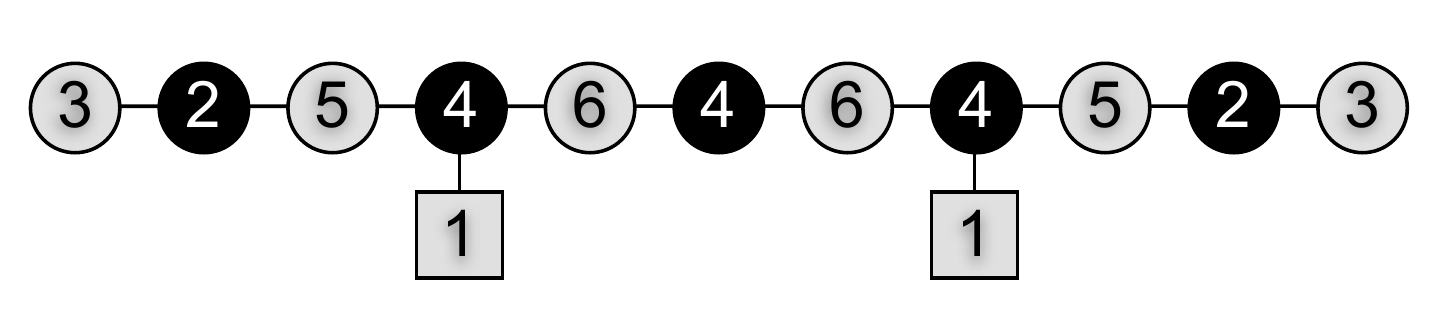}\qquad \includegraphics[width=5cm]{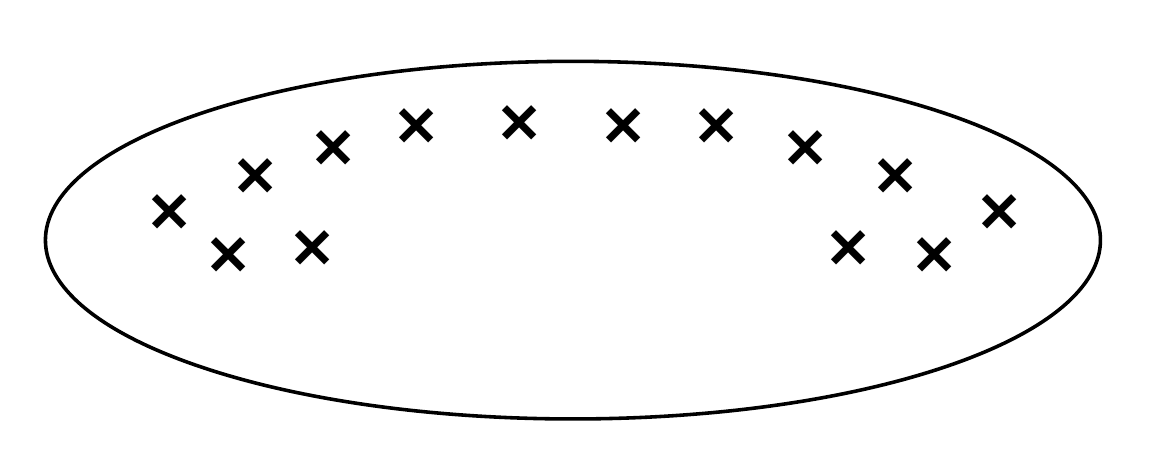}
\\[5mm]
\includegraphics[width=4.5cm]{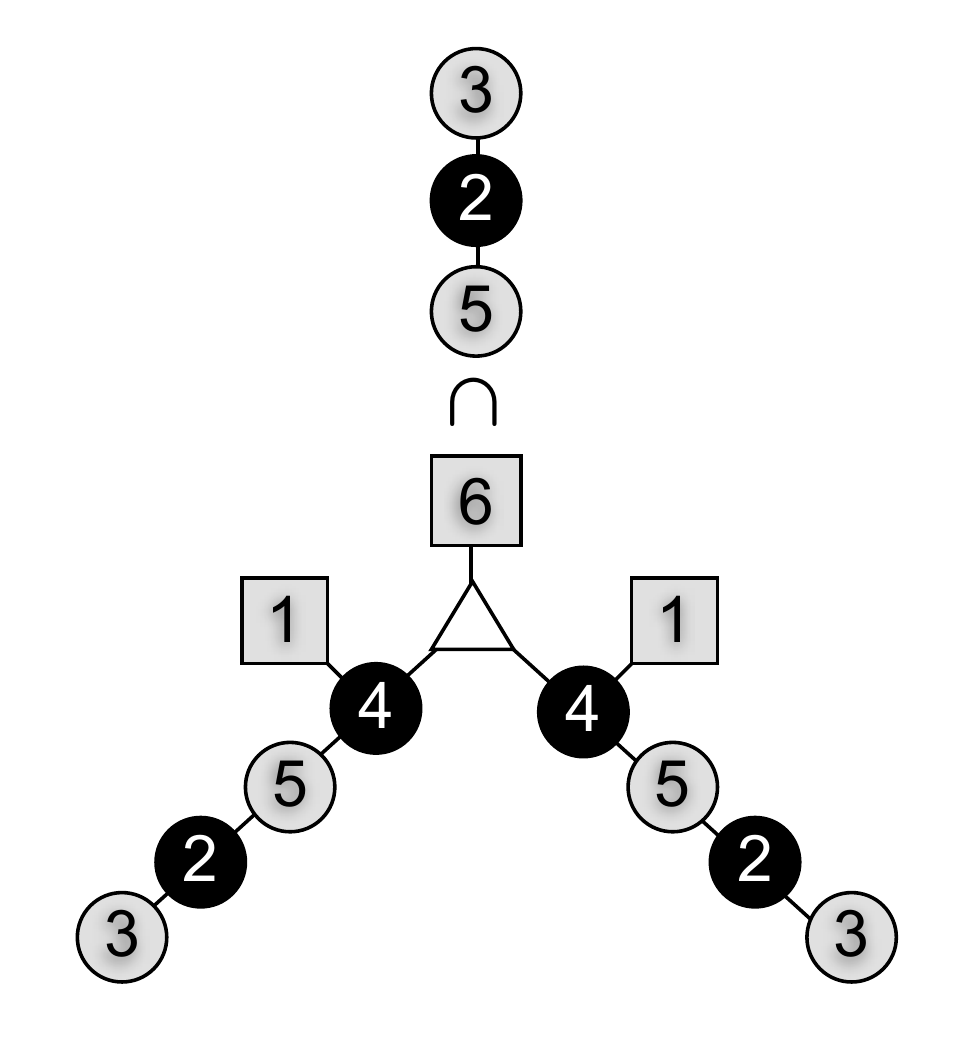}\qquad\qquad\qquad\includegraphics[width=4.5cm]{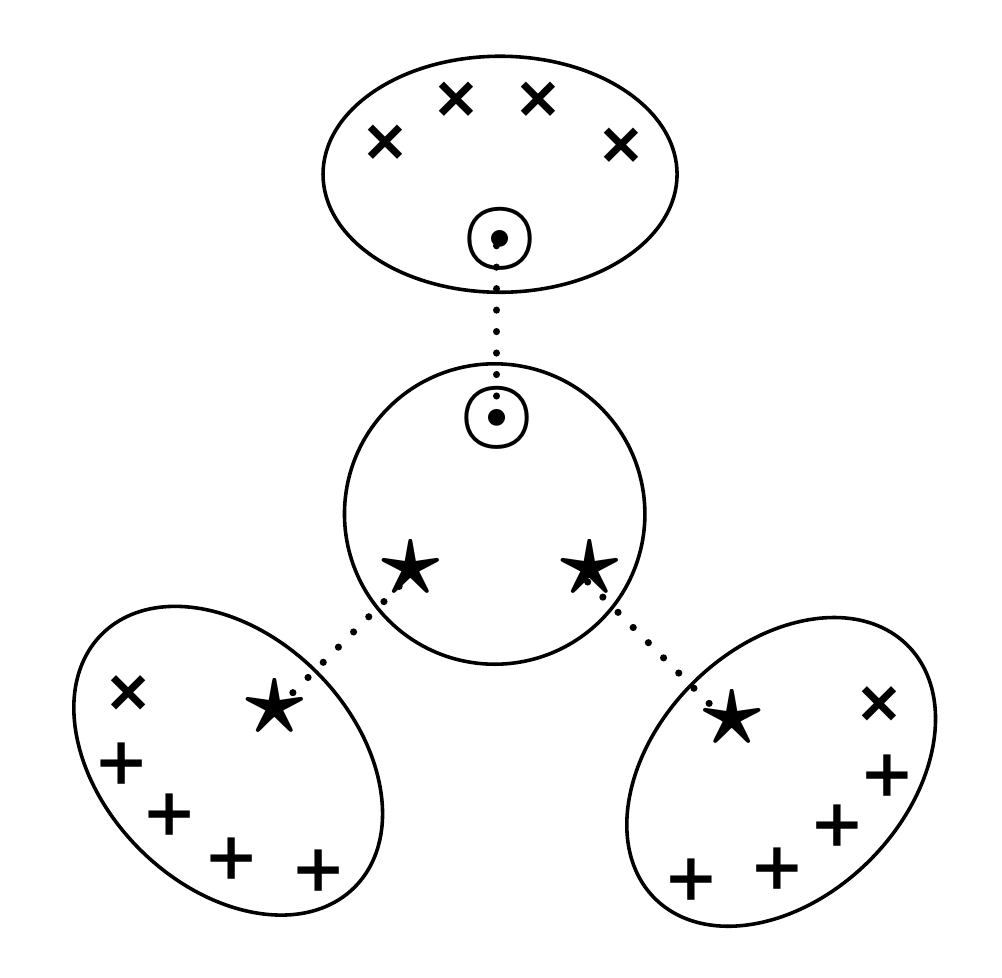}
\caption{The construction of $\widetilde{T}_{SO(2N)}$ from a linear quiver in the $N=3$ case.}
\label{fig:T_USp}
\end{center}
\end{figure}

We can also consider a theory with $SO(2N)\times USp(2N-2)^2$ flavor symmetry without any marginal coupling, which we denote by $\widetilde{T}_{SO(2N)}$. Such a theory is constructed by considering a linear quiver of $6N-7$ gauge groups
\begin{eqnarray}
&& SO(3) \times USp(2) \times \cdots SO(2N-1)\times
\nonumber\\
&&\qquad \qquad USp(2N-2) \times SO(2N) \times \cdots SO(2N)\times USp(2N-2)\times
\nonumber\\
&&\qquad\qquad\qquad\qquad SO(2N-1)\times USp(2N-4) \times \cdots \times USp(2)\times SO(3), \qquad
\label{eq:6N-7}
\end{eqnarray}
whose G-curve is a sphere with $6N-4$ punctures of $\times$ (See figure \ref{fig:T_USp}). We now deform this curve so that it has a single tail with $2N-2$ punctures of $\times$ and two tails with $2N-1$ punctures of $\times$. 
The former tail describes the the linear quiver with gauge groups \eqref{eq:tail_SO}. On the other hand, each of the latter tails involves a linear quiver with gauge groups
\begin{eqnarray}
 USp(2N-2)\times SO(2N-1)\times USp(2N-4) \times \cdots \times USp(2) \times SO(3).
\end{eqnarray}
The three tails are connected to a sphere with a puncture of $\odot$ and two punctures of $\star$, which implies that the sphere describes $\widetilde{T}_{SO(2N)}$ theory. We can calculate $n_v$ and $n_h$ for this $\widetilde{T}_{SO(2N)}$ theory, exactly in the same way as for $T_{SO(2N)}$. The result is
\begin{eqnarray}
 n_{v}(\widetilde{T}_{SO(2N)}) = \frac{8N^3}{3} -7N^2 + \frac{16N}{3} -1,\qquad n_{h}(\widetilde{T}_{SO(2N)}) = \frac{8N^3}{3} - 4N^2 + \frac{4N}{3}.
\label{eq:anomaly_TSO2}
\end{eqnarray}
Note that there is no similar theory with $USp(2N-2)^3$ or $SO(2N)^2 \times USp(2N-2)$ flavor symmetry because the puncture $\star$ has a non-trivial $\mathbb{Z}_2$ monodromy around it \cite{Tachikawa:2009rb}.

Furthermore, we can also consider a Riemann surface of genus $g>0$ as a G-curve. Then, the low energy gauge theory is described by $2N$ M5-branes wrapping on the Riemann surface of genus $g$, which is not S-dual to any linear quiver theory. In general, theories which are related to each other by S-duality have the same punctures and genus of the G-curve. In the next section, we will consider the gravity dual of $SO/USp$ quivers whose G-curve has no puncture. The dual gravity for various $SO/USp$-type tails is studied in section \ref{sec:gravity2}.

\section{The gravity dual without punctures}
\label{sec:gravity1}

In this section, we consider the gravity duals of the $SO/USp$ superconformal quivers whose G-curve is a Riemann surface of genus $g$ without punctures. We first review the gravity duals of $SU$ quivers in \ref{subsec:SU}, which was studied in \cite{Gaiotto:2009gz}, and then generalize it to the $SO/USp$ quivers in \ref{subsec:SO/USp}. The main difference from $SU$ quivers are the $\mathbb{Z}_2$-quotient in the bulk. We holographically calculate the conformal anomaly of the four-dimensional theory, which agrees with the calculation in the field theory side.

What is interesting here is that in general various different quiver gauge theories are associated with the same G-curve, which is quite different from the $SU$-type quivers studied in \cite{Gaiotto:2009we, Gaiotto:2009gz}. In subsection \ref{subsec:SO/USp2}, we show that such theories associated with the same G-curve have the same conformal anomalies, which suggests that their gravity duals share the same metric. Then, in subsection \ref{subsec:torsion}, we discuss that the gravity duals of such theories are further classified by the torsion part of the four-form flux.

\subsection{The gravity dual of $SU(M)$ quivers}
\label{subsec:SU}

The gravity duals of $SU(M)$ quivers that arise as the low energy theories of $M$ M5-branes on a Riemann surface was studied in \cite{Gaiotto:2009gz}. In particular, when the Riemann surface has genus $g>1$ and no puncture, the near-horizon geometry of the M5-branes is identified with $AdS_5\times \Sigma_g\times S^4$, where $S^4$ is non-trivially fibered over the Riemann surface $\Sigma_g$. The corresponding eleven-dimensional metric is given by
\begin{eqnarray}
ds^2 &=& (\pi Ml_p^3)^{\frac{2}{3}}\frac{W^{\frac{1}{3}}}{2}\left\{4ds^2_{AdS_5} + 2\left[4\frac{dr^2 + r^2 d\beta^2}{(1-r^2)^2}\right] + 2d\theta^2 \right.
\nonumber\\
&& \left.+ \frac{2}{W}\cos^2\theta (d\psi^2 + \sin^2\psi d\phi^2) + \frac{4}{W}\sin^2\theta \left(d\chi + \frac{2r^2d\beta}{1-r^2}\right)^2
\right\},
\label{eq:no-puncture}
\end{eqnarray}
where $W \equiv 1 + \cos^2\theta$. The parameters $\theta, \psi,\phi$ and $\chi$ are the coordinates on $S^4$, while $r$ and $\beta$ parameterize a hyperbolic space. We obtain a Riemann surface by considering a quotient of the hyperbolic space by some group $\Gamma$. This metric has a symmetry $SU(2) \times U(1)$ which is suitable for the gravity dual of $d=4,\mathcal{N}=2$ theories. From the above metric, the conformal anomaly $c$ of the four-dimensional gauge theory is holographically evaluated as \cite{Henningson:1998gx, Gaiotto:2009gz}
\begin{eqnarray}
c = \frac{\pi R_{AdS_5}^3}{8G_{\rm N}^5} = \frac{\pi R_{AdS_5}^3\times {\rm Vol}(\Sigma_g\times S^4)}{8G_{\rm N}^{11}} = \frac{M^3}{3}(g-1),
\label{eq:gravity_c}
\end{eqnarray}
where $G_{\rm N}^d$ is $d$-dimensional Newtonian constant and $R_{AdS_5}$ is the curvature radius of the $AdS_5$.

\begin{figure}
\begin{center}
\includegraphics[width=13cm]{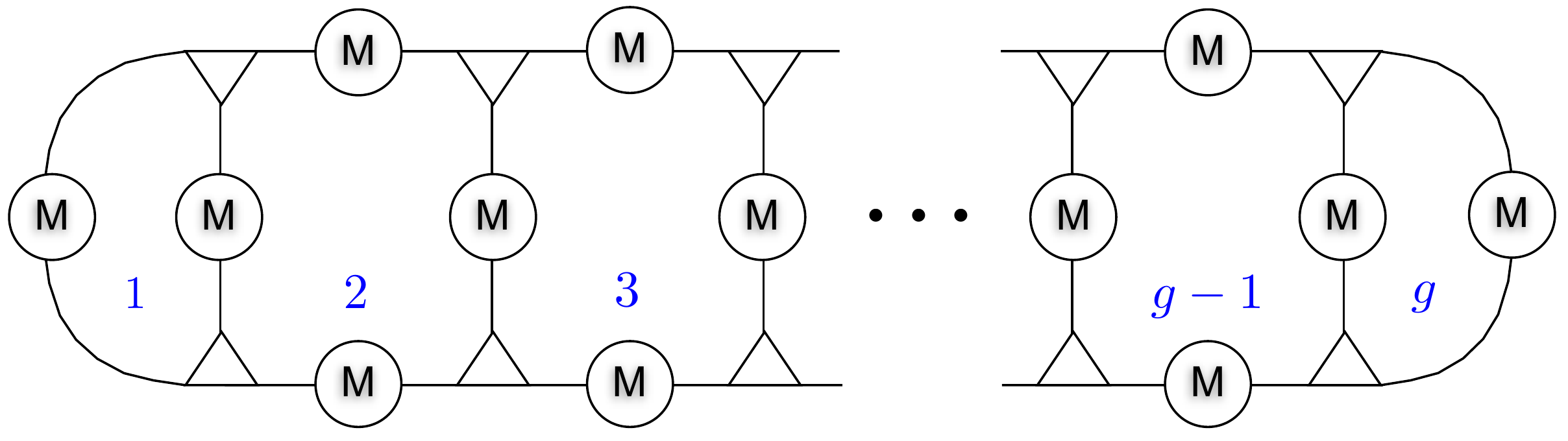}
\caption{The quiver diagram of the theory whose G-curve is a Riemann surface of genus $g$ without punctures. Each circle represents $SU(N)$ gauge group while each white triangle expresses the $T_M$ theory.}
\label{fig:genus_g}
\end{center}
\end{figure}

The result \eqref{eq:gravity_c} in the gravity side matches with the field theory calculation. The field theory whose G-curve is a Riemann surface of genus $g$ without punctures is described by a quiver diagram as in figure \ref{fig:genus_g}, which involves mysterious $T_{M}$ theories with $SU(M)^3$ flavor symmetry and no marginal coupling \cite{Gaiotto:2009we}.
The anomaly contributions $n_v$ and $n_h$ of $T_{M}$ theory can be read off via S-duality invariance \cite{Gaiotto:2009gz}, which leads to the following total contributions from the whole quiver: 
\begin{eqnarray}
 n_{v}({\rm total}) = (g-1)\left(\frac{4M^3}{3}-\frac{M}{3}-1\right),\qquad n_{h}({\rm total}) = (g-1)\left(\frac{4M^3}{3}-\frac{4M}{3}\right).
\label{eq:nvnh_SU}
\end{eqnarray}
 Then the total anomaly coefficient $c$ is evaluated in the field theory side as
\begin{eqnarray}
 c = \frac{2n_{v}({\rm total})+n_{h}({\rm total})}{12} = \left[\frac{M^3}{3} -\frac{M}{6} -\frac{1}{6}\right](g-1),
\end{eqnarray}
which agrees with the gravity side calculation \eqref{eq:gravity_c} in the large $M$ limit. In \cite{Gaiotto:2009gz}, it was also shown that subleading contributions in \eqref{eq:nvnh_SU} are also consistent with the anomaly polynomial of the six-dimensional $(2,0)$ theory which was calculated by considering the anomaly inflow from the bulk \cite{Harvey:1998bx}.

\subsection{$\mathbb{Z}_2$-quotient in the gravity dual}
\label{subsec:SO/USp}

We now generalize the above argument to the $SO/USp$ superconformal quivers. The main difference from the $SU$-type quivers is the presence of a $\mathbb{Z}_2$-quotient. Namely, we consider $2N$ M5-branes on top of $\mathbb{R}^5/\mathbb{Z}_2$ which are wrapping on a Riemann surface of genus $g$ without punctures.
From table \ref{table:O4-1} and table \ref{table:O4-2}, we find that such a configuration gives D4-branes with O4$^\pm$-plane in the type IIA limit, which lead to $SO(2N),USp(2N-2)$ gauge groups in four dimensions. The $\mathbb{Z}_2$-quotient only affects $\mathbb{R}^5$ spanned by $x^{4,5,7,8,9}$, which replaces the $S^4$ surrounding M5-branes with $\mathbb{RP}^4$. Then, the corresponding gravity dual is described by the same metric \eqref{eq:no-puncture} with $M=2N$ but now $\theta,\psi,\phi$ and $\chi$ parameterize $\mathbb{RP}^4$ rather than $S^4$.

\vspace*{5mm}
\noindent{\em Anomaly matching}
\vspace*{3mm}

\begin{figure}
\begin{center}
\includegraphics[width=13cm]{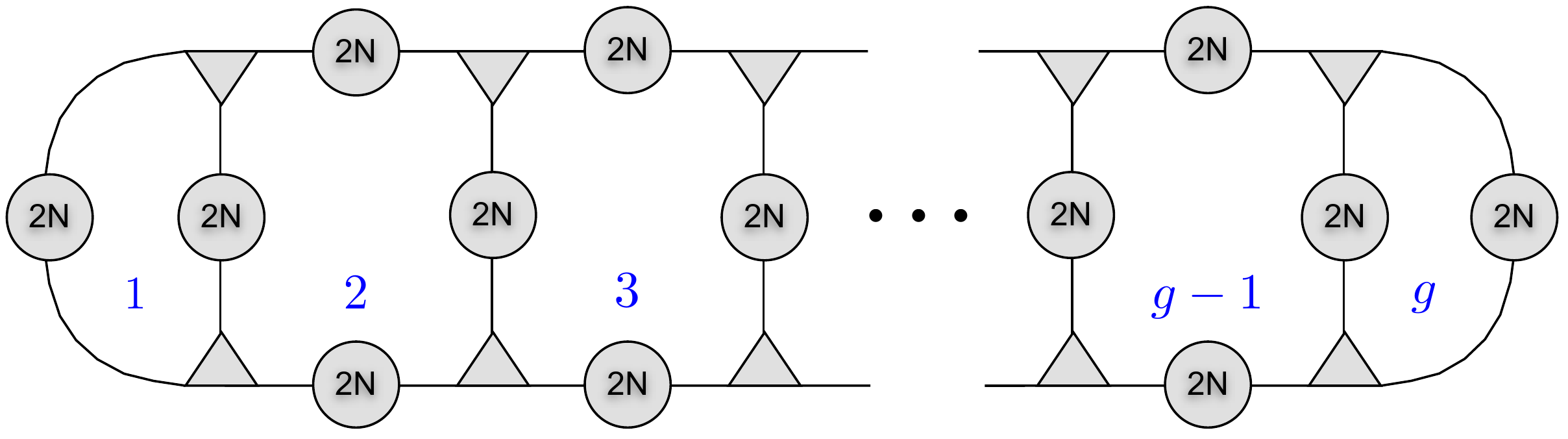}
\caption{A $SO/USp$ superconformal quiver gauge theory whose G-curve is a Riemann surface of genus $g$ without punctures. A grey circle represents $SO(2N)$ gauge group, while a grey triangle expresses $T_{SO(2N)}$ theory.}
\label{fig:no-puncture_SO}
\end{center}
\end{figure}

As a $SO/USp$ superconformal quiver gauge theory whose G-curve is a Riemann surface of genus $g$ without punctures, we can consider a theory described by a quiver diagram depicted in figure \ref{fig:no-puncture_SO}. 
It involves $(3g-3)$ $SO(2N)$ gauge groups and $(2g-2)$ $T_{SO(2N)}$ theories, and therefore the total contributions to $n_v$ and $n_h$ are
\begin{eqnarray}
 n_{v}({\rm total}) &=& (g-1)\left[\frac{16N^3}{3} - 8N^2 + \frac{11N}{3}\right],
\label{eq:nvnh_SO/USp1} \\
n_{h}({\rm total}) &=& (g-1)\left[\frac{16N^3}{3} - 8N^2 + \frac{8N}{3}\right].
\label{eq:nvnh_SO/USp2}
\end{eqnarray}
Then the conformal anomaly $c$ of the quiver gauge theory is evaluated as
\begin{eqnarray}
 c = \frac{2n_{v}({\rm total}) + n_{h}({\rm total})}{12} = (g-1)\left[\frac{4N^3}{3} - 2N^2 + \frac{5N}{6}\right].
\label{eq:c_SO}
\end{eqnarray}
On the other hand, the holographic calculation from the gravity dual gives
\begin{eqnarray}
 c= \frac{\pi R^3_{AdS_5}}{8G_{\rm N}^5} = \frac{\pi R^3_{AdS_5}\times {\rm Vol}(\Sigma_g\times \mathbb{RP}^4)}{8G_{\rm N}^{11}} = \frac{4N^3}{3}(g-1),
\label{eq:gravity_SO/USp}
\end{eqnarray}
which agrees with \eqref{eq:c_SO} in the large $N$ limit. This supports the validity of \eqref{eq:anomaly_TSO1} in the large $N$ limit, which was calculated via the S-duality invariance of the four-dimensional field theory.

\vspace*{5mm}
\noindent{\em Finite $N$ corrections}
\vspace*{3mm}

For completeness, we here describe that the subleading terms in \eqref{eq:nvnh_SO/USp1} and \eqref{eq:nvnh_SO/USp2} are consistent with the anomaly polynomial of the six-dimensional field theory. A field theory on $2N$ M5-branes on top of $\mathbb{R}^5/\mathbb{Z}_2$ is the $D_N$-type $(2,0)$ theory. The anomaly polynomial of the theory is evaluated in \cite{Yi:2001bz} so that the anomaly coming from the polynomial is canceled by the anomaly inflow through the Chern-Simons coupling in eleven-dimensional gravity. The explicit expression for the anomaly eight-form is
\begin{eqnarray}
 I_8 = N\mathcal{J}_8 + N(2N-1)(2N-2)\frac{p_2(\mathcal{N})}{24},
\label{eq:anomaly8}
\end{eqnarray}
where $p_i(\mathcal{B})$ is the $i$-th Pontryagin class of a bundle $\mathcal{B}$, and $\mathcal{N}$ is the normal bundle of the M5-brane world-volume. The eight-form $\mathcal{J}_8$ is the one-loop anomaly polynomial of a single $(2,0)$ tensor multiplet:
\begin{eqnarray}
 \mathcal{J}_8 = \frac{1}{48}\left[p_2(\mathcal{N}) - p_2(\mathcal{T}) + \frac{(p_1(\mathcal{T}) - p_1(\mathcal{N}))^2}{4}\right],
\end{eqnarray}
where $\mathcal{T}$ is the tangent bundle of the world-volume.

By integrating this $I_8$ along the Riemann surface, we can obtain the anomaly six-form $I_6$ of the four-dimensional theory. The anomaly coefficients of $I_6$ give R-symmetry anomalies in four dimensions, which are related to the conformal anomalies via superconformal symmetry. Therefore, we can read off $a$ and $c$ from the anomaly coefficients of $I_6$. In fact, such a calculation was carried out in the appendix of \cite{Alday:2009qq}, which tells us that the resulting conformal anomalies are written as
\begin{eqnarray}
 a = (g-1)\frac{5N + 8N(2N-1)(2N-2)}{24},\quad c = (g-1)\frac{N+2N(2N-1)(2N-2)}{6}.
\nonumber\\
\end{eqnarray}
By using \eqref{eq:ac-nvnh}, we can then read off $n_v$ and $n_h$ of the quiver gauge theory, which perfectly agree with \eqref{eq:nvnh_SO/USp1} and \eqref{eq:nvnh_SO/USp2}.\footnote{For some interesting observations on the anomaly polynomials of the $(2,0)$ theory in related topics, see \cite{Benini:2009mz, Bonelli:2009zp, Nishioka:2011jk}.}

\subsection{Inclusion of $\widetilde{T}_{SO(2N)}$ theories}
\label{subsec:SO/USp2}

\begin{figure}
\begin{center}
\includegraphics[width=4cm]{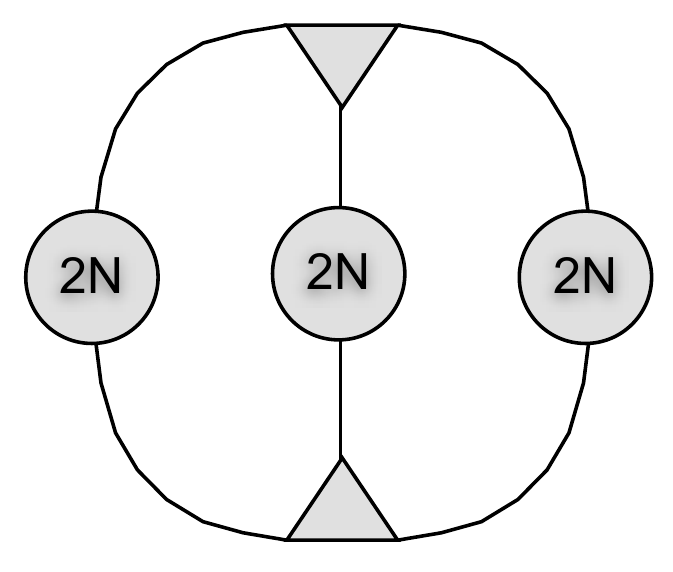}\qquad\qquad \includegraphics[width=4cm]{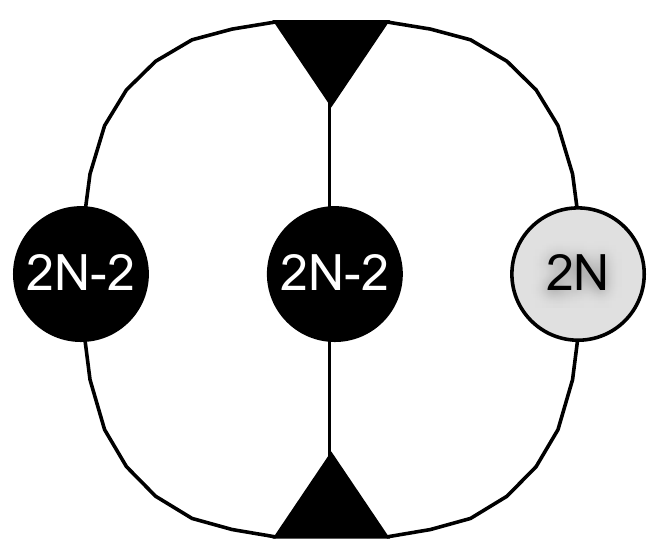}
\caption{Two quiver gauge theories whose G-curve is $\Sigma_2$. A black circle represents $USp(2N-2)$ gauge group, and a black triangle expresses $\widetilde{T}_{SO(2N)}$ theory. The left quiver involves three $SO(2N)$ gauge groups and two $T_{SO(2N)}$ theories, while the right one includes two $USp(2N-2)$ and a single $SO(2N)$ gauge groups as well as two $\widetilde{T}_{SO(2N)}$ theories.}
\label{fig:g=2_USp}
\end{center}
\end{figure}

We have other four-dimensional quiver gauge theories whose G-curve is the same $\Sigma_g$. For example, let us consider the simplest case $g=2$. In this case, we have previously considered the theory with three $SO(2N)$ gauge groups and two $T_{SO(2N)}$ theories. However, we can also consider a theory with two $USp(2N-2)$ and a single $SO(2N)$ gauge groups together with two $\widetilde{T}_{SO(2N)}$ theories, whose quiver diagram is shown in the right picture of figure \ref{fig:g=2_USp}.

The latter theory has two $USp(2N-2)$ gauge groups instead of $SO(2N)$. Recall that the M-theory lifts of $SO(2N)$ and $USp(2N-2)$ gauge theories for $\varphi=0$ are both given by $2N$ M5-branes on $\mathbb{R}^5\times \mathbb{R}^5/\mathbb{Z}_2\times S^1$. The only difference is that two of $2N$ M5-branes are localized at the $\mathbb{Z}_2$ fixed plane for $USp(2N-2)$ gauge theory. This ``freezing'' of two M5-branes is due to the non-vanishing torsion part of the four-form flux \cite{Hori:1998iv}.

Since the torsion element of the four-form flux can be expressed by a flat three-form potential, we expect that such a torsion part does not affect the anomaly eight-form \eqref{eq:anomaly8} evaluated by the anomaly inflow method. If this is the case, the two theories in figure \ref{fig:g=2_USp} have the same conformal anomalies. In order to verify this, we calculate $n_v$ and $n_h$ for the right quiver of figure \ref{fig:g=2_USp}. By using the expressions \eqref{eq:anomaly_TSO2} for the anomalies of the $\widetilde{T}_{SO(2N)}$, the total contributions to $n_v$ and $n_h$ are evaluated as
\begin{eqnarray}
n_v({\rm total}) = \frac{16N^3}{3}-8N^2 + \frac{11N}{3},\qquad n_h({\rm total}) = \frac{16N^3}{3} - 8N^2 + \frac{8N}{3},
\label{eq:g=1_USp}
\end{eqnarray}
which are exactly the same as \eqref{eq:nvnh_SO/USp1} and \eqref{eq:nvnh_SO/USp2} for $g=2$. This strongly suggests that the two quiver gauge theories associated with the two quiver diagrams in figure \ref{fig:g=2_USp} share the same metric of the dual gravity as studied in the previous subsection.

\vspace*{5mm}
\noindent{\em General Genus case}
\vspace*{3mm}

\begin{figure}
\begin{center}
\includegraphics[width=15cm]{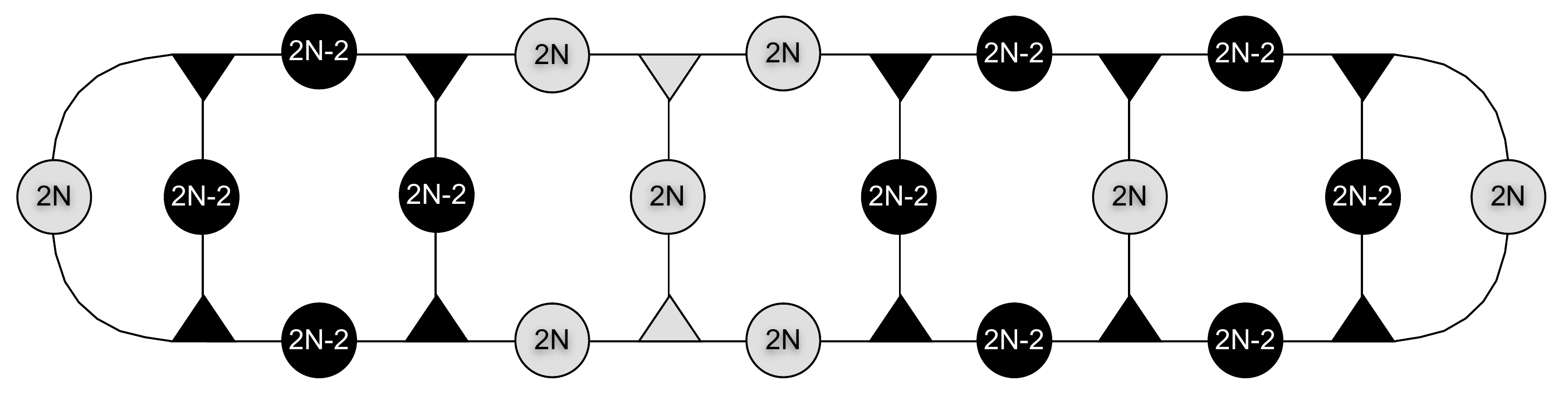}
\caption{An example of superconformal quivers constructed by $SO(2N), USp(2N-2)$ gauge groups and $T_{SO(2N)}, \widetilde{T}_{SO(2N)}$ theories. This example is for $g=7$.}
\label{fig:no-puncture_USp}
\end{center}
\end{figure}

We can generalize the above argument on the conformal anomalies to theories with $\Sigma_g$ for $g>2$. Let us consider an arbitrary superconformal which is constructed from $SO(2N)$ and $USp(2N-2)$ gauge groups, $T_{SO(2N)}$ theories, and $\widetilde{T}_{SO(2N)}$ theories. An example is shown in figure \ref{fig:no-puncture_USp}.
Any such quiver can be obtained by replacing some $SO(2N)$ and $T_{SO(2N)}$ with $USp(2N-2)$ and $\widetilde{T}_{SO(2N)}$ in a quiver of figure \ref{fig:no-puncture_SO}. Since the $\widetilde{T}_{SO(2N)}$ theory has $SO(2N)\times USp(2N-2)^2$ flavor symmetry, such a replacement should be realized by repeating primitive replacements defined below.

The primitive replacement is defined as a replacement of a closed chain of $SO(2N)$ gauge groups and $T_{SO(2N)}$ theories of the form
\sbox{\boxa}{\includegraphics[width=15cm]{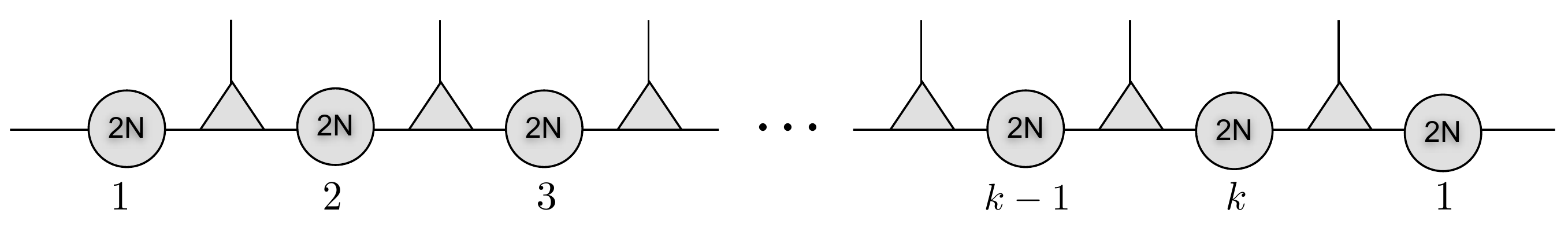}}
\settowidth{\bw}{\usebox{\boxa}}
\begin{eqnarray}
\parbox{\bw}{\usebox{\boxa}},\nonumber\\
\end{eqnarray}
with the following closed chain of $USp(2N-2)$ gauge groups and $\widetilde{T}_{SO(2N)}$ theories:
\sbox{\boxa}{\includegraphics[width=15cm]{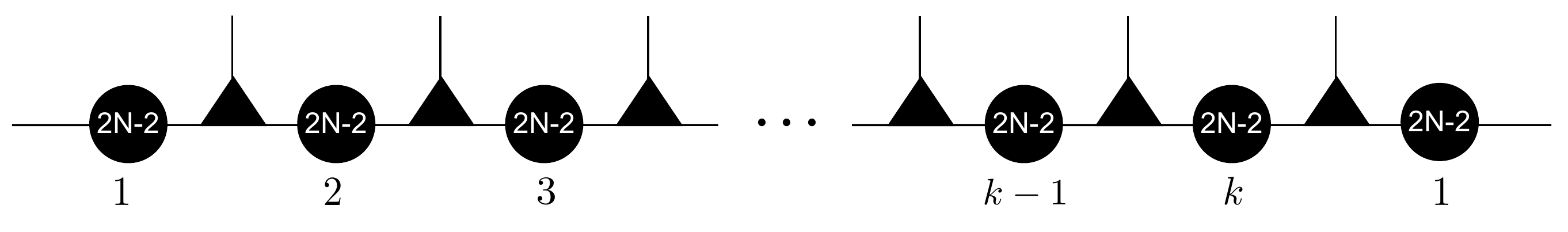}}
\settowidth{\bw}{\usebox{\boxa}}
\begin{eqnarray}
\parbox{\bw}{\usebox{\boxa}}.\nonumber\\
\end{eqnarray}
Here the leftmost and rightmost gauge groups in each chain are identified.
Since the original chain has equal numbers, say $k$, of $SO(2N)$ gauge groups and $T_{\rm SO(2N)}$ theories, the primitive replacement replaces $k$ $SO(2N)$ with $k$ $USp(2N-2)$ as well as $k$ $T_{SO(2N)}$ with $k$ $\widetilde{T}_{SO(2N)}$. What is important here is that this replacement keeps $n_v$ and $n_h$ invariant, which follows from
\begin{eqnarray}
 n_v(SO(2N)) + n_v(T_{SO(2N)}) &=& n_v(USp(2N-2)) + n_v(\widetilde{T}_{SO(2N)}),
\\
 n_h(T_{SO(2N)}) &=& n_h(\widetilde{T}_{SO(2N)}).
\end{eqnarray}

Since any quiver which only involves $SO/USp$ gauge groups and $T_{SO},\widetilde{T}_{SO}$ theories can be obtained by repeating the primitive replacements in a quiver of figure \ref{fig:no-puncture_SO}, it has the same $n_v$ and $n_h$ as \eqref{eq:nvnh_SO/USp1} and \eqref{eq:nvnh_SO/USp2}. Hence, if a $SO/USp$ superconformal quiver has a G-curve $\Sigma_g$ without punctures, then its conformal anomalies are determined only by the genus $g$ of the G-curve and independent of the choice of gauge groups associated with the handles of $\Sigma_g$. This particularly suggests that they share the same metric of the dual gravity as the one studied in subsection \ref{subsec:SO/USp}. In the next subsection, we see how we can distinguish the gravity duals of such theories with the same G-curve $\Sigma_g$.

\subsection{The torsion part of four-form flux}
\label{subsec:torsion}

We have seen that there are several quiver gauge theories which have the same G-curve $\Sigma_g$ and the same anomaly contributions $n_v, n_h$. In fact, such theories are further classified by the torsion part of the four-form flux in M-theory.\footnote{The author thanks Yuji Tachikawa for pointing out this fact.}

Let us first consider M5-branes on $\mathbb{R}^5\times \mathbb{R}^5/\mathbb{Z}_2 \times S^1$. Since the spacetime (from which the M5-brane locus is removed) is contractible to $\mathbb{RP}^4\times S^1$, the topology of the three-form potential is measured by $H^4(\mathbb{RP}^4\times S^1,\widetilde{\mathbb{Z}})$.\footnote{Note here that, since the $\mathbb{Z}_2$-fixed plane carries M5 charge $-1$ (counted in the covering space), the four-form flux $[G_4/2\pi]$ itself is not a cohomology class in $H^4(\mathbb{RP}^4\times S^1,\widetilde{\mathbb{Z}})$ \cite{Hori:1998iv}. To make an element of $H^4(\mathbb{RP}^4\times S^1,\widetilde{\mathbb{Z}})$, we need to define a modified cohomology class
$[\widetilde{G}_4/2\pi] \equiv \frac{1}{2}([G_4/\pi]-\chi_{\rm E})\in H^4(\mathbb{RP}^4\times S^1,\widetilde{\mathbb{Z}})$,
where $\chi_{\rm E}$ denotes the twisted Euler class. Then, the $H$-flux in type IIA limit is obtained by $\int_{S^1}[\widetilde{G}_4/2\pi]= [H/2\pi] \in H^3(\mathbb{RP}^4,\widetilde{\mathbb{Z}})$. For more detail, see the appendix of \cite{Hori:1998iv}.} As discussed in \cite{Hori:1998iv}, the cohomology group $H^4(\mathbb{RP}^4\times S^1, \widetilde{\mathbb{Z}})$ includes two-torsion:
\begin{eqnarray}
 H^4(\mathbb{RP}^4\times S^1,\widetilde{\mathbb{Z}}) \;\simeq\; \mathbb{Z} \oplus \mathbb{Z}_2,
\label{eq:torsion}
\end{eqnarray}
where the integrations over $\mathbb{RP}^4$ and $S^1$ induce projections $H^4(\mathbb{RP}^4 \times S^1,\widetilde{\mathbb{Z}})\to \mathbb{Z}$ and $H^4(\mathbb{RP}^4\times S^1,\widetilde{\mathbb{Z}})\to \mathbb{Z}_2$, respectively. This means that the first $\mathbb{Z}$ in \eqref{eq:torsion} counts the number of M5-branes wrapping on $\mathbb{R}^5\times S^1$, while the torsion part $\mathbb{Z}_2$ can be identified with $H^3(\mathbb{RP}^4,\widetilde{\mathbb{Z}})$ which measures the topology of NSNS $B$-field in the type IIA limit. In particular, the phase $\vartheta$ defined in \eqref{eq:topology} is equivalent to the integral of the four-form flux along $S^1\times \mathbb{RP}^3$.\footnote{Since the flux is a twisted four-form and $\mathbb{RP}^3\times S^1$ is an {\em untwisted} cycle, this integral is well-defined as an element of $\mathbb{Z}_2$.} Thus, the difference between O4$^+$ and O4$^-$-planes is lifted to a difference in the torsion part $\mathbb{Z}_2$.

\begin{figure}
\begin{center}
\includegraphics[width=4.8cm]{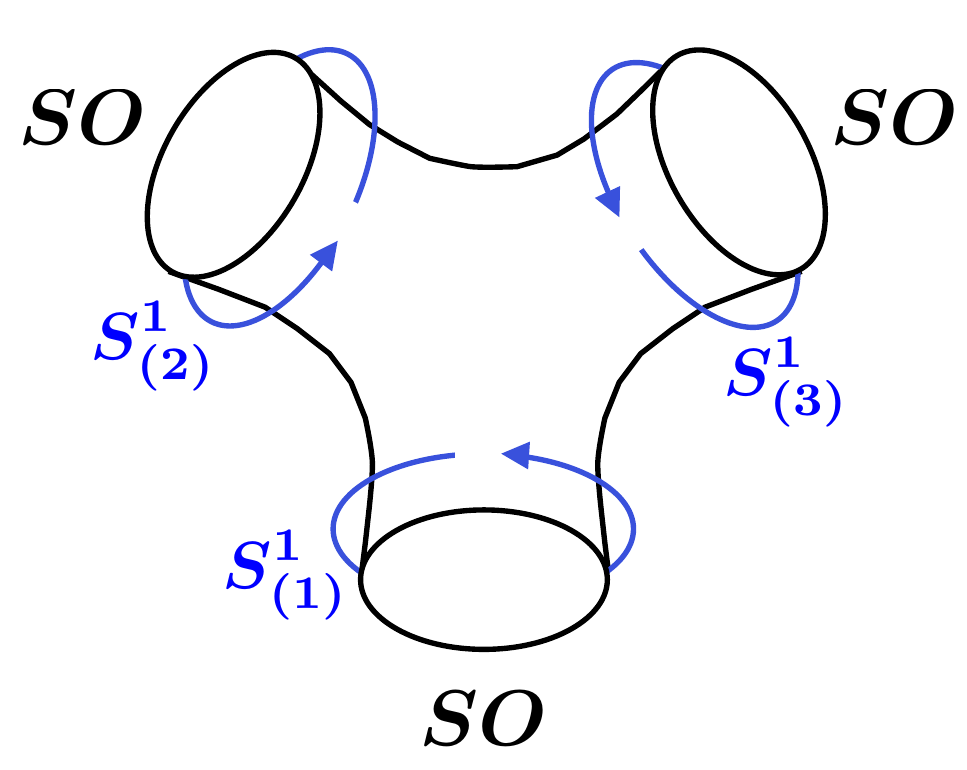}\qquad\qquad
\includegraphics[width=5cm]{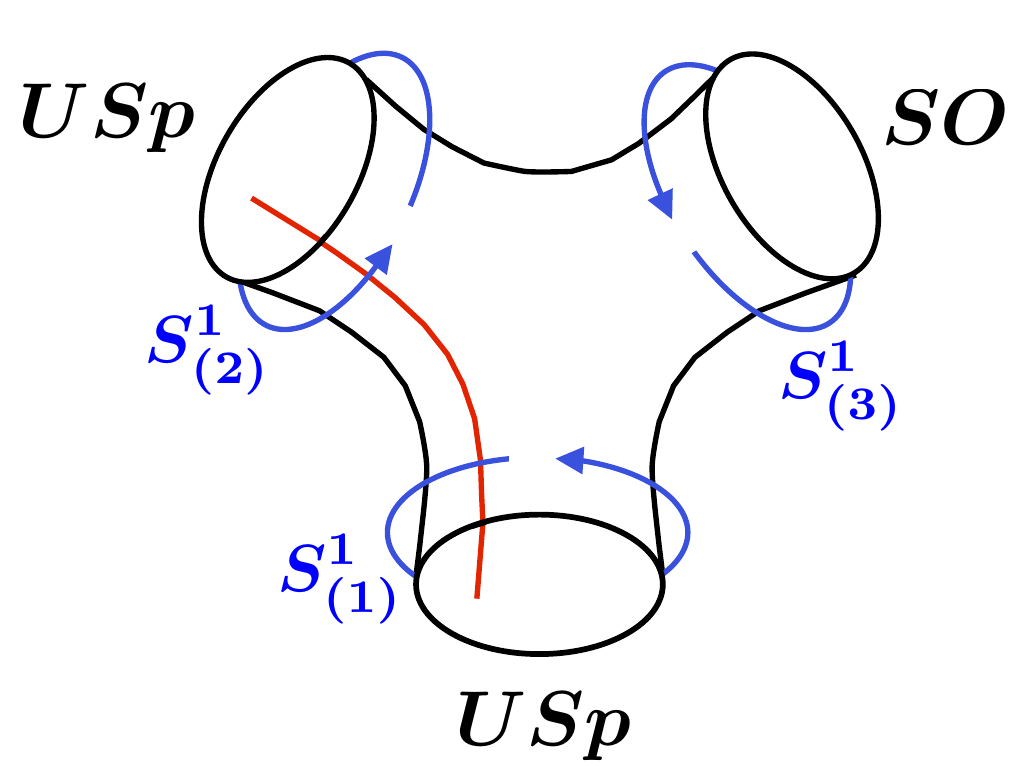}
\caption{Left: The ``pair of pants'' in a G-curve associated with $T_{SO}$-theory. \,\,Right: Its $\widetilde{T}_{SO}$-theory counterpart. We can draw a red line along which the non-vanishing torsion of the four-form flux is turned on.}
\label{fig:T_SO-torsion}
\end{center}
\end{figure}

Now, let us consider two types of ``pairs of pants'' in the G-curve as in figure \ref{fig:T_SO-torsion}, which correspond to the M-theory lifts of $T_{SO}$ and $\widetilde{T}_{SO}$-theories, respectively. 
Each type of the pants has three tubes attached to it, and each such tube is associated with a one-cycle $S^1_{(i)}$ for $i=1,2,3$ with the orientation as in figure \ref{fig:T_SO-torsion}. The torsion part of the four-form flux associated to the $i$-th tube is measured by integrating the flux along $S^1_{(i)}\times \mathbb{RP}^3$. We denote such integral of the flux by $\vartheta_{i}$ for the $i$-th tube.\footnote{For the same reason as before, this flux is the modified flux $[\widetilde{G}_4/2\pi]$. To be more specific, $\vartheta_i = \int_{S^1_{(i)}\times \mathbb{RP}^4}[\widetilde{G}_4/2\pi]$.} The left and right examples in figure \ref{fig:T_SO-torsion} have $(\vartheta_1,\vartheta_2,\vartheta_3) = (0,0,0)$ and $(1,1,0)$, respectively. Since $\vartheta_i\in \mathbb{Z}_2$, one might think that a single pair of pants generally has four possibilities of $(\vartheta_1,\vartheta_2,\vartheta_3)$ up to permutation. However, the fact $\sum_{i}S^1_{(i)} = 0$ implies that 
\begin{eqnarray}
\sum_{i=1}^3 \vartheta_i \;=\; 0
\end{eqnarray}
as an element of $\mathbb{Z}_2$, which forbids $(\vartheta_1,\vartheta_2,\vartheta_3) = (1,0,0),\,(1,1,1)$. This is consistent with the absence of the four-dimensional theory which has $SO(2N)^2\times USp(2N-2)$ or $USp(2N-2)^3$ flavor symmetry and no marginal coupling.

\begin{figure}
\begin{center}
\includegraphics[width=13cm]{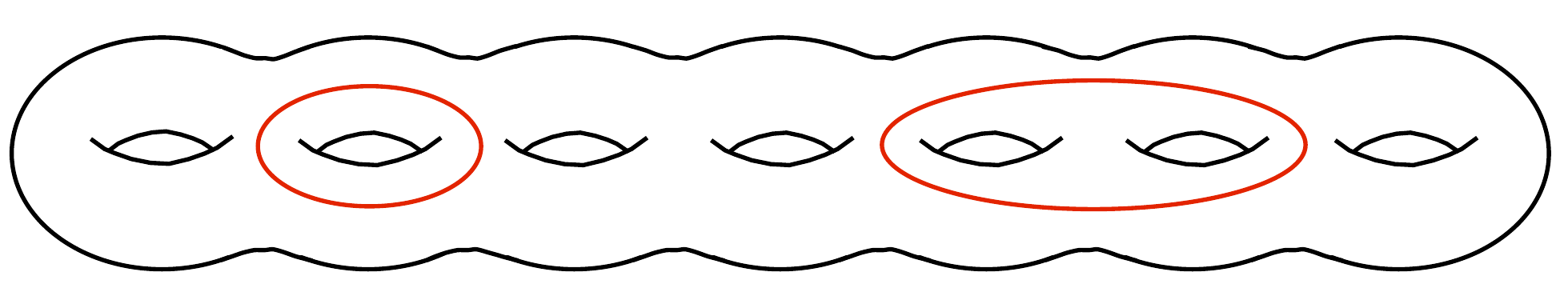}
\caption{The non-vanishing torsion part of the four-form flux is turned on along the ``$B$-cycles'' of the G-curve. This is the example for the theory of figure \ref{fig:no-puncture_USp}.}
\label{fig:torsion-g}
\end{center}
\end{figure}

The non-trivial torsion part of the four-form flux can be expressed by drawing a red line through tubes with $\vartheta_i\neq 0$, as in figure \ref{fig:T_SO-torsion}. Due to the absence of the pair of pants which has $SO(2N)^2\times USp(2N-2)$ or $USp(2N-2)^3$ flavor symmetry, such a red line should form a closed curve (figure \ref{fig:torsion-g}). This particularly implies that the non-vanishing torsion part of the four-form flux is associated to one-cycles of the G-curve which are transverse to the M-theory circle, as long as the curve has no punctures.

It is now clear that the gravity duals of $SO/USp$ quivers with the same G-curve $\Sigma_g$ are further classified by the torsion part of the four-form flux which is associated to the ``$B$-cycles'' of the G-curve. The example for the quiver of figure \ref{fig:no-puncture_USp} is shown in figure \ref{fig:torsion-g}.
In other words, the gravity duals of $SO/USp$ quivers whose G-curve has no punctures are fully classified by the genus $g$ of the G-curve and the torsion part of the four-form flux.

\section{Solutions for $SO/USp$ tails}
\label{sec:gravity2}

We here discuss the gravity duals of various $SO/USp$ punctures on the G-curve. The punctures are constructed by inserting some additional M5-branes on the $2N$ M5-branes studied in the previous section. We need to find its appropriate gravity dual which is consistent with the $\mathbb{Z}_2$-quotient. Without the $\mathbb{Z}_2$-quotient, such a $M5$-brane insertion rather gives $SU$-type punctures studied in \cite{Gaiotto:2009gz}. We briefly review the dual gravity of the $SU$-type tails in \ref{subsec:SU-puncture}, and consider its $\mathbb{Z}_2$-quotient in \ref{subsec:SO-puncture}.

\subsection{$SU$-type tails}
\label{subsec:SU-puncture}

\begin{figure}
\begin{center}
\includegraphics[width=7cm]{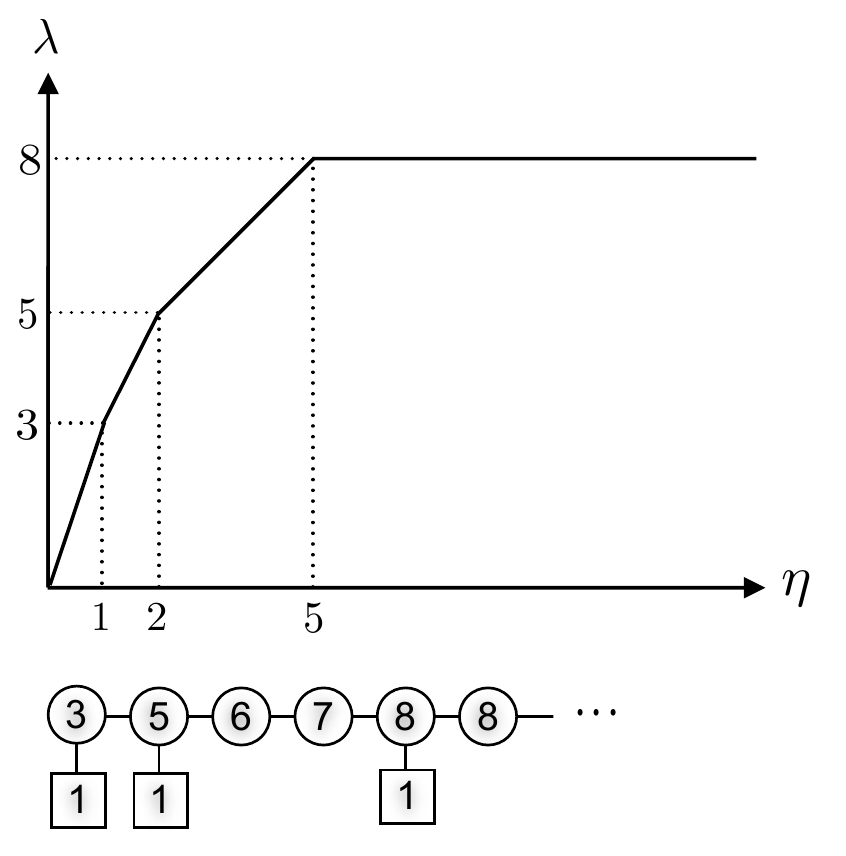}
\caption{An example of the line charge density $\lambda(\eta)$ for a tail of $SU(8)$ quivers. In the $SU$-type quiver, the number in a white circle represents the rank of the gauge group, while that in a white box expresses the number of fundamentals.}
\label{fig:line-charge}
\end{center}
\end{figure}

The dual gravity solutions for $SU$-type punctures were studied in \cite{Gaiotto:2009gz} by using the general construction of half-BPS solutions in eleven-dimensional supergravity. In particular, a solution which is $U(1)$-symmetric around the puncture is obtained by solving an axially symmetric electrostatics problem in three dimensions:
\begin{eqnarray}
\ddot{V} + \rho^2 V'' = 0.
\label{eq:Laplace}
\end{eqnarray}
Here, we used the short-hand notations $\dot{V}= \rho \frac{\partial V}{\partial \rho}$ and $V'=\frac{\partial V}{\partial \eta}$, where $\rho$ is the radial coordinate of a two-dimensional plane and $\eta$ is the ``height'' coordinate parameterizing the third direction. The equation \eqref{eq:Laplace} has a solution with a line charge density at $\rho = 0$:
\begin{eqnarray}
 \dot{V}|_{\rho = 0} = \lambda(\eta).
\end{eqnarray}
For each $SU$-type tail, this line charge density $\lambda(\eta)$ is determined uniquely as follows. We consider the region $\eta\geq 0$, and first determine $\lambda(i)=N_i$ for $i\in \mathbb{N}$ so that the $i$-th gauge group from the edge of the tail is $SU(N_i)$ (figure \ref{fig:line-charge}).
The value of $\lambda(\eta)$ for $\eta\not\in\mathbb{N}$ is determined so that $\lambda(\eta)$ has a constant slope in the interval $(i,i+1)$. Then, the eleven-dimensional metric of the gravity dual near the puncture is written in terms of $V$ associated with the boundary condition $\dot{V}|_{\rho=0}=\lambda(\eta)$. In particular, in the vicinity of $\rho=0$, the metric and the 3-form potential are written as
\begin{eqnarray}
ds^2_{11} &\sim& \kappa^{\frac{2}{3}}\left(\frac{\dot{V}\widetilde{\Delta}}{2V''}\right)^{\frac{1}{3}}\left[4ds^2_{AdS_5} + \frac{2V''\dot{V}}{\widetilde{\Delta}}ds^2_{S^2} + \frac{2V''}{\dot{V}}\left(d\rho^2 + \rho^2d\chi^2 + d\eta^2\right) \right.
\nonumber\\
&&\left. + \frac{4}{\widetilde{\Delta}}\left(d\beta + \dot{V}'d\chi\right)^2\right],
\label{eq:metric-puncture}\\
 \widetilde{\Delta} &\sim& 2\dot{V}V'' + (\dot{V}')^2,
\nonumber\\
C_3 &\sim& \frac{1}{8\pi^2}\left[(-\dot{V}+\eta \dot{V}')d\chi + \left(\frac{\dot{V}\dot{V}'}{\widetilde{\Delta}} - \eta\right)(d\beta + \dot{V}'d\chi)\right]d\Omega_2,
\label{eq:three-form}
\end{eqnarray}
where $\beta$ and $\chi$ have period $2\pi$.
Since we are now considering the region $\rho\sim 0$, we can use the approximation $\dot{V}' \sim \lambda'(\eta)$. The various checks of this solution were performed in \cite{Gaiotto:2009gz, Tachikawa:2011dz}.

\subsection{$\mathbb{Z}_2$-quotient for tails}
\label{subsec:SO-puncture}

Now, we consider a generalization of the above solutions to the $SO/USp$-type tails. To identify the dual gravity, we need to determine $\lambda(\eta)$ for $SO/USp$-type tails and take into account the $\mathbb{Z}_2$-quotient in the bulk.

\begin{figure}[h]
\begin{center}
\includegraphics[width=6.9cm]{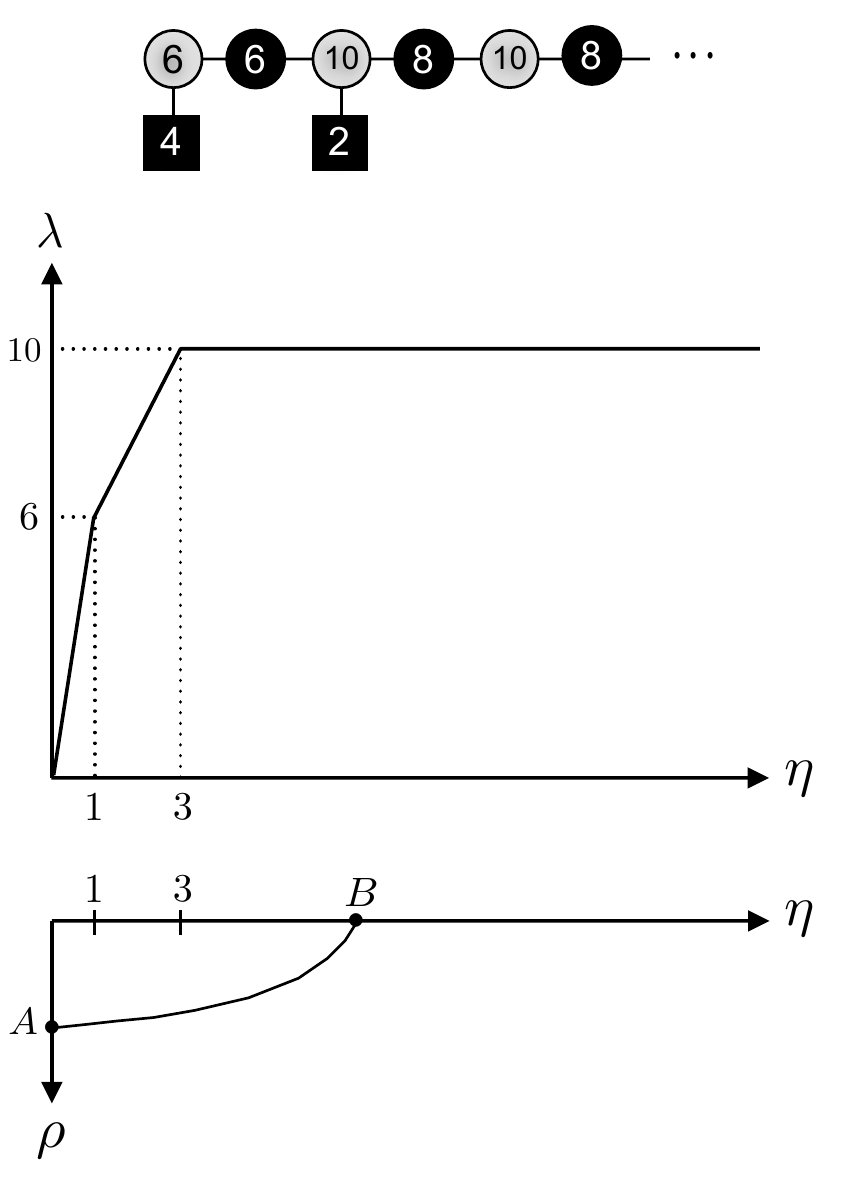}\quad
\includegraphics[width=7.3cm]{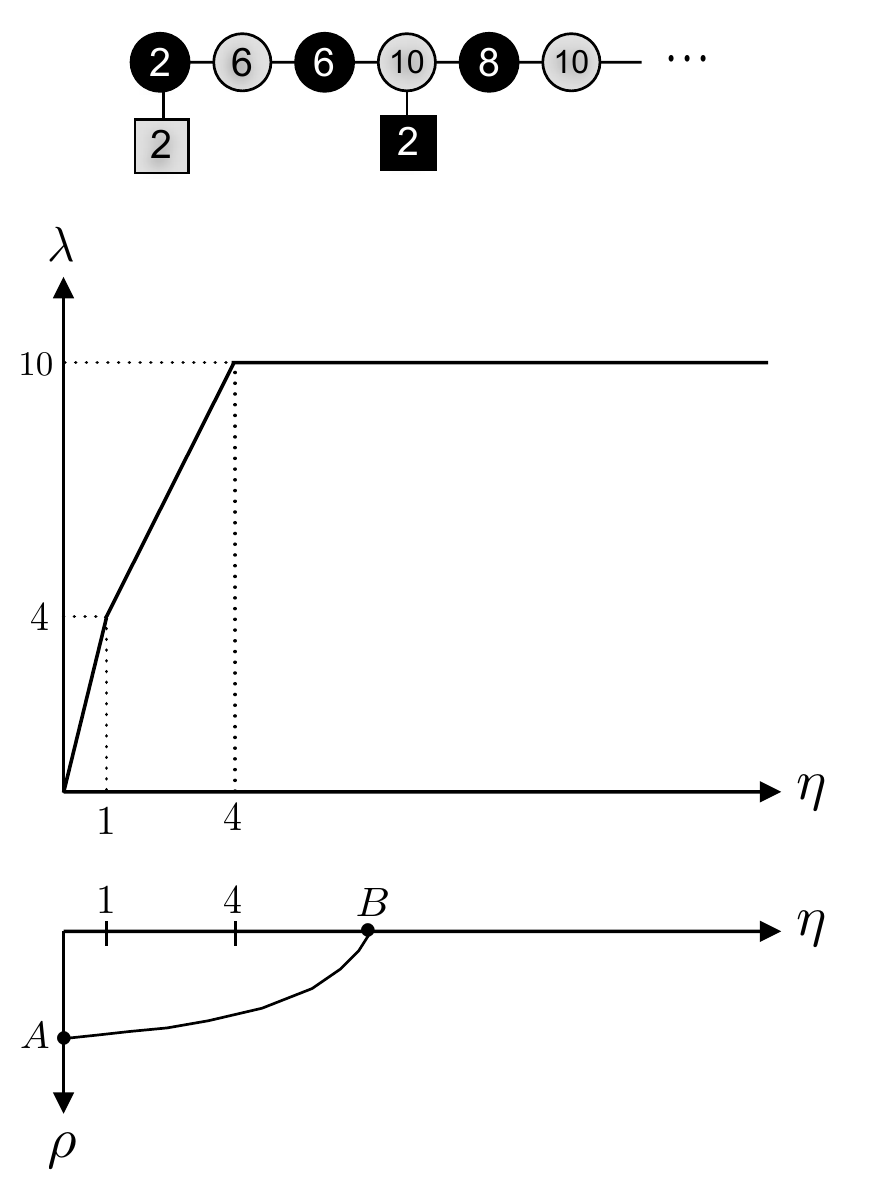}
\caption{Two examples of the line charge densities $\lambda(\eta)$ for $SO/USp$ tails. The bottom pictures shows the line segment $\overline{AB}$ mentioned in the main text. The $S^1_{\chi} \times S^2$ bundle over the segment is topologically $S^4$, which is surrounding $2N$ M5-branes wrapping on the Riemann surface.}
\label{fig:SO-tail}
\end{center}
\end{figure}

To identify $\lambda(\eta)$ for a $SO/USp$ tail, we first note that in the metric \eqref{eq:metric-puncture} the space spanned by $\rho,\chi,\eta$ and $\beta$ has a structure which is similar to the multi Taub-NUT space. In fact, since the slope $\dot{V}'|_{\rho=0} = \lambda'(\eta)$ can change only at $\eta=i\in \mathbb{N}$, we can generally write $\dot{V}''|_{\rho=0} = -\sum_{i} k_i\delta(\eta-i)$ with $k_i\geq 0$. This means that, in the vicinity of points $(\rho,\eta) = (0,i)$ for $k_i\neq 0$, the quantity $V''$ can be approximately written as
\begin{eqnarray}
 V'' \sim \frac{k_i}{2}\frac{1}{\sqrt{\rho^2+(\eta-i)^2}}.
\end{eqnarray}
Thus, we find that the four-dimensional space we are considering has $\mathbb{C}^2/\mathbb{Z}_{k_i}$ singularity near $\eta=i$, which corresponds to $k_i$ D6-branes in the type IIA configuration.\footnote{Away from the points $(\rho,\eta)=(0,i)$, the four-dimensional metric differs from that of Taub-NUT space, due to the backreaction of M5-branes.} On the other hand, since the number $k_i$ is equivalent to the change of the slope $\lambda'(\eta)$ at $\eta=i$, we have a relation
\begin{eqnarray}
 k_i = 2\lambda(i)-\lambda(i-1) - \lambda(i+1).
\end{eqnarray}
By comparing this with equation \eqref{eq:superconformal_SO/USp}, we find that 
\begin{eqnarray}
\lambda(i) = d_i,
\end{eqnarray}
for $i=1,2,3,\cdots$.
We determine the value of $\lambda(\eta)$ for $i<\eta<i+1$ so that it has a constant slope in the interval $(i,i+1)$. Two examples are shown in figure \ref{fig:SO-tail}.

What we need to do next is to identify the $\mathbb{Z}_2$-quotient in the bulk. From table \ref{table:O4-1} and table \ref{table:O4-2}, we can see that the near horizon geometry of M5-branes which give $SO/USp$ gauge theory in four dimensions involves $S^1 \times \mathbb{RP}^4$ or $(S^1 \times S^4)/\mathbb{Z}_2$, depending on the value of $\varphi$. The phase $\varphi$ measures a non-trivial RR $U(1)$ gauge field background in the type IIA setup. In particular, crossing a D6-brane shifts $\varphi$ by one unit, changing the near horizon geometry.
We need to find an appropriate $\mathbb{Z}_2$-quotient in the bulk which is consistent with this property. 

For that, we first note that a four-cycle which surrounds the M5-branes wrapped on the Riemann surface is constructed as follows \cite{Gaiotto:2009gz}. Let us consider a line segment $\overline{AB}$ in $(\rho, \eta)$-plane as depicted in figure \ref{fig:SO-tail}, which starts at $\eta=0,\rho\neq 0$ and ends at $\rho=0,\eta=\eta_1$ for $\eta_1$ satisfying $\lambda(\eta_1) = 2N$. We consider a $S^1\times S^2$ fibration over the line segment, where the $S^2$ is that for the second term in the metric \eqref{eq:metric-puncture} and shrinks into zero size at the point $A$. On the other hand, the $S^1$ is parameterized by $\chi$ and shrinks into zero at the point $B$. Thus, the bundle we are considering is topologically $S^4$. From the explicit expression for $C_3$ in \eqref{eq:metric-puncture}, it follows that the total four-form flux through this four-cycle is $2N$, which implies that the four-cycle is surrounding the $2N$ M5-branes wrapped on the Riemann surface.

Since such M5-branes are those studied in the previous section, the $\mathbb{Z}_2$-quotient should only affect the four-cycle surrounding the $2N$ M5-branes, leaving the Riemann surface invariant. To be more specific, the $\mathbb{Z}_2$-quotient just replaces the four-cycle with $\mathbb{RP}^4$, which is equivalent to the replacement
\begin{eqnarray}
 S^1_\chi\to S^1_\chi\,/\,\mathbb{Z}_2, \qquad S^2\to \mathbb{RP}^2,
\label{eq:Z2_puncture}
\end{eqnarray}
where the $\mathbb{Z}_2$-action on $S^1_\chi$ is a half-period shift. Note that the $\mathbb{Z}_2$ trivially acts on other coordinates in the metric \eqref{eq:metric-puncture}. Then, the resulting geometry is described by
\begin{eqnarray}
 ds^2_{11} &\sim& \kappa^{\frac{2}{3}}\left(\frac{\dot{V}\widetilde{\Delta}}{2V''}\right)^{\frac{1}{3}}\left[4ds^2_{AdS_5} + \frac{2V''\dot{V}}{\widetilde{\Delta}}ds^2_{\mathbb{RP}^2} + \frac{2V''}{\dot{V}}(d\rho^2 + \rho^2d\chi^2 + d\eta^2)\right.
\nonumber\\
&&\left.+ \frac{4}{\widetilde{\Delta}}\left(d\beta + \dot{V}'d\chi\right)^2\right],
\label{eq:metric-puncture2}
\end{eqnarray}
with identifications $\beta \sim \beta +2\pi$ and $\chi \sim \chi+\pi$.

\begin{figure}
\begin{center}
\includegraphics[width=10cm]{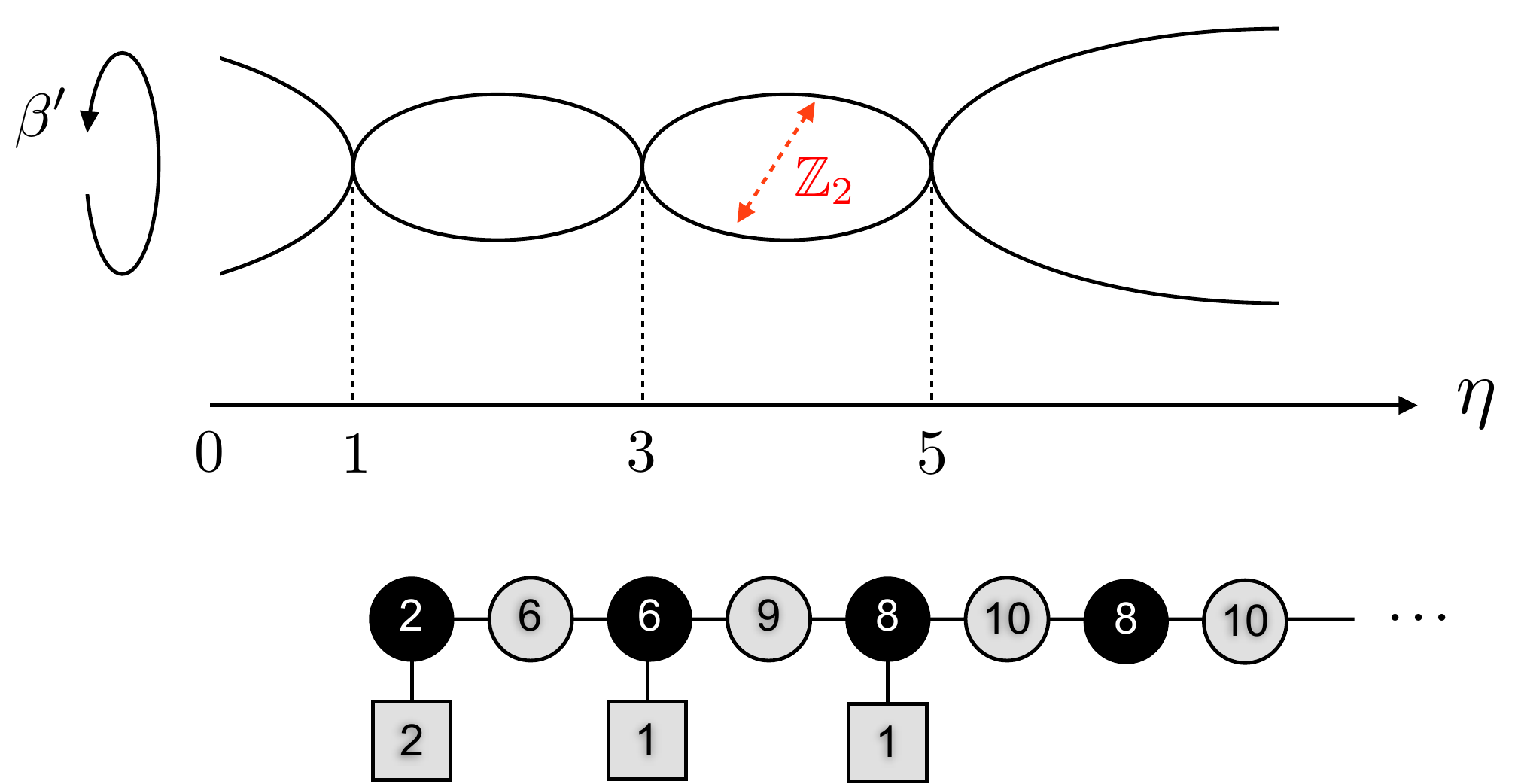}
\caption{The $\mathbb{Z}_2$-action on $S^1_{\beta'}$. In the example, the right-most two-cycle is the ``middle'' $\mathbb{P}^1$, and the left tail has three points where $S^1_{\beta'}$ degenerates into point. When one crosses the points $\eta = 3,5$, the non-trivial $\mathbb{Z}_2$-action appears or disappears because the slope $\dot{V}'|_{\rho=0}$ changes by one. On the other hand, crossing $\eta=1$ does not changes the $\mathbb{Z}_2$-action on $S^1_{\beta'}$. The reason for this is that at the point $\eta=1$ the slope $\dot{V}'|_{\rho=0}$ changes by an even number.}
\label{fig:Taub-NUT-like}
\end{center}
\end{figure}

Now, we verify that the $\mathbb{Z}_2$-quotient \eqref{eq:Z2_puncture} is consistent with the previously mentioned property of the $SO/USp$-type tails. Recall that, before the $\mathbb{Z}_2$-quotient, the four-dimensional space spanned by $\rho,\chi,\eta$ and $\beta$ has a structure similar to the multi Taub-NUT space. In particular, there is a chain of two-cycles in the subspace $\rho=0$. Each two-cycle can be regarded as a $S^1$-fibration over a line segment on $\eta$-axis (figure \ref{fig:Taub-NUT-like}).
Here the $S^1$ is parameterized by
\begin{eqnarray}
\beta' \equiv \beta + \dot{V}'\chi,
\label{eq:beta'}
\end{eqnarray}
which shrinks into zero size at points where the slope $\dot{V}'$ discontinuously changes.
Then, how does the $\mathbb{Z}_2$-quotient \eqref{eq:Z2_puncture} affect these two-cycles? The replacement $S^2\to \mathbb{RP}^2$ does not affect the four-dimensional space we are considering. On the other hand, the replacement $S^1_\chi\to S^1_\chi/\mathbb{Z}_2$ changes the period of $\chi$-direction into $\chi\sim \chi +\pi$. This does or does not affect $S^1_{\beta'}$, depending on the value of $\dot{V}'$. In fact, the $\mathbb{Z}_2$-quotient implies an identification
\begin{eqnarray}
 (\chi,\beta') \sim (\chi+\pi,\,\beta'+\dot{V}'\pi),
\end{eqnarray}
in $(\chi,\beta')$-plane, which is trivial for $S^1_{\beta'}$ if $\dot{V}'$ is even, while it is non-trivial for $S^1_{\beta'}$  if $\dot{V}'$ is odd. In other words, the non-trivial $\mathbb{Z}_2$-action on $S^1_{\beta'}$ appears or disappears when $\dot{V}'$ changes by one, that is, when a single D6-brane is crossed in the type IIA configuration. This is in perfect agreement with the property of $SO/USp$-type tails, where the non-trivial $\mathbb{Z}_2$-action on the Taub-NUT circle appears or disappears when one crosses a Taub-NUT center, as depicted in figure \ref{fig:Taub-NUT}.

\section{Discussions}

In this paper, we have studied the gravity dual solutions of $SO/USp$ superconformal quiver gauge theories which are realized by the IR limits of M5-branes on a Riemann surface together with $\mathbb{Z}_2$-quotient. In section \ref{sec:gravity1}, we have considered the gravity duals of the theories whose G-curve is a Riemann surface without punctures. The dual geometry is determined by the genus $g$ of the G-curve, and holographically gives the correct four-dimensional conformal anomalies. We have also found that there are generally several theories which have the same G-curve $\Sigma_g$ and the same conformal anomalies $n_v$ and $n_h$. The gravity duals of such theories share the same metric of the near horizon geometry, but are further classified by the torsion part of the four-form flux associated to the ``$B$-cycles'' of the G-curve. In section \ref{sec:gravity2}, we have considered the gravity duals of the $SO/USp$-type tails. We have identified the correct line charge density $\lambda(\eta)$ and $\mathbb{Z}_2$-quotient in the bulk, which is consistent with the property of the $SO/USp$-tails.

For future direction, it would be interesting to study the gravity duals of $SO/USp$ quivers whose G-curve has genus $g>1$ and various punctures. For that, we need to solve the Toda equation rather than the axially symmetric electrostatics problem \cite{Gaiotto:2009gz}, and also take into account the torsion part of the four-form flux. The existence of the torsion part will give a rich class of gravity duals of $d=4,\mathcal{N}=2$ superconformal theories.

It would also be interesting to study the relation between the torsion part of the four-form flux and the outer-automorphism twist on the G-curve. As pointed out in \cite{Tachikawa:2009rb}, some of the $SO/USp$ punctures have a $\mathbb{Z}_2$-monodromy around them which flips the sign of a world-volume scalar field on the G-curve. This is related to the outer-automorphism twist on the G-curve \cite{Tachikawa:2010vg}, which distinguishes $SO$ and $USp$ gauge groups associated with the handles of the curve. On the other hand, as explained in section \ref{sec:gravity1}, we can distinguish the two gauge groups by the torsion part of the four-form flux, if the $SO$ and $USp$ gauge groups are realized by O4$^\pm$-planes. However, if the two gauge groups are realized by O4$^0$ or $\widetilde{\rm O4}^+$-plane, then there is no torsion element of the four-form flux \cite{Hori:1998iv}. In fact, the difference between O4$^0$ and $\widetilde{\rm O4}^+$ is lifted to the difference between even and odd elements of $H^4((S^4\times S^1)/\mathbb{Z}_2,\widetilde{\mathbb{Z}})\simeq \mathbb{Z}$.\footnote{The integration along $S^1$ maps the even and odd elements of $H^4((S^4\times S^1)/\mathbb{Z}_2,\widetilde{\mathbb{Z}})$ to zero and non-zero elements in $H^3(\mathbb{RP}^4,\widetilde{\mathbb{Z}})\simeq \mathbb{Z}_2$, respectively.} In terms of the line charge density $\lambda(\eta)$, this difference corresponds to whether $\lambda(i)$ is even or odd for $i\in \mathbb{N}$. It would be interesting to perform further study on the relation between this and the outer-automorphism twist.

\subsection*{Acknowledgments} 
The author thanks Yuji Tachikawa for illuminating discussions, important comments and careful reading of this manuscript. This work is partially supported by the National Research Foundation of Korea (NRF) grant funded by the Korea government (MEST) 2005-0049409.

\end{document}